%% file: ms.tex
\documentclass[12pt,preprint,eqsecnum]{aastex}
\usepackage{epsfig}
\usepackage{natbib}
\bibpunct{(}{)}{;}{a}{}{,}

\input{commands}

\shorttitle{Cen~A middle lobe as a buoyant bubble}
\shortauthors{Saxton et al.}

\begin{document}

\title{The Centaurus~A Northern Middle Lobe as a Buoyant Bubble}
\author{Curtis~J.~Saxton\altaffilmark{1,2}
\and {Ralph~S.~Sutherland\altaffilmark{2}}
\and {Geoffrey~V.~Bicknell\altaffilmark{2,1}}
}

\altaffiltext{1}{Department of Physics \& Theoretical Physics,
	Faculty of Science,
	Australian National University, ACT 0200, Australia }
\altaffiltext{2}{Research School of Astronomy \& Astrophysics,
	Mt Stromlo Observatory, Australian National University,
	Cotter Road, Weston ACT 2611, Australia }




\begin{abstract}
We model the northern middle radio lobe of Centaurus~A (NGC~5128)
as a buoyant bubble of plasma
deposited by an intermittently active jet.
The extent of the rise of the bubble and its morphology imply
that the ratio of its density to that of the surrounding ISM
is less than $10^{-2}$,
consistent with our knowledge of extragalactic jets
and minimal entrainment into the precursor radio lobe.
Using the morphology of the lobe
to date the beginning of its rise through the atmosphere of Centaurus A,
we conclude that the bubble has been rising for approximately 140~Myr.
This time scale is consistent with that proposed by
Quillen et al. (1993)
for the settling of post-merger gas into the 
presently observed large scale disk in NGC~5128,
suggesting a strong connection between the delayed 
re-establishment of radio emission and the
merger of NGC~5128 with a small gas-rich galaxy.
This suggests a connection,
for radio galaxies in general,
between mergers and the {\em delayed} onset of radio emission.
In our model, the elongated X-ray emission region discovered by
Feigelson et al. (1981),
part of which coincides with the northern middle lobe,
is thermal gas that originates from the ISM below the bubble
and that has been uplifted and compressed.
The ``large-scale jet'' appearing in the radio images of Morganti et al. (1999) 
may be the result of the same pressure gradients
that cause the uplift of the thermal gas, acting on much lighter plasma,
or may represent a jet that did not turn off completely
when the northern middle lobe started to buoyantly rise.
We propose that the adjacent emission line knots (the ``outer filaments'')
and star-forming regions result from the disturbance,
in particular the thermal trunk,
caused by the bubble moving through the extended atmosphere of NGC~5128.
\end{abstract}


\newpage

\section{INTRODUCTION}

Centaurus A,
as the closest active galaxy that is also radio loud
has always attracted a lot of attention
with numerous papers covering the electromagnetic spectrum
from radio through to gamma-rays.
Its proximity means that various physical processes can be studied in
detail with high spatial resolution
and from time to time new observational data appear that
confront our understanding of this object
and the physical processes occurring in radio galaxies generally.
The recent observations of the northern middle lobe (NML) of Centaurus~A
by \citet{morganti1999}
and the related observations of star formation
and emission-line clouds in its vicinity
\citep{graham1981, morganti1991, graham1998, sutherland1993, mould}
fall into this category.
The radio observations of the NML
are the first published observations of the NML
and fill in the spatial gap between the extant VLA images of the inner lobe
and the large scale structure.
The curious structure of the NML
and its associated ``large-scale jet'' linking it the inner lobe
beg explanation as does the apparent association with
the star-forming and emission-line regions.
We propose such an explanation in this paper
and in so doing link the formation of multi-lobed structure in Centaurus~A
to its aspherical X-ray morphology and its merger history.
Our work is based in the growing recognition of the importance
of buoyant effects in the structure of radio galaxies.
\citet{churazov} have presented a convincing model of the M87 radio structures
and associated X-ray features
proposing that the intermediate radio structures in that galaxy \citep{owen}
can be modeled in terms of radio lobes rising buoyantly
in the gravitational potential.
In this paper we propose a similar model for Centaurus~A
albeit with some differences that are related to
different aspects of Centaurus~A.
In particular,
we concentrate on the timescale and history of the bubble's rise,
and its possible relationship with the merger history of NGC~5128
\citep{tubbs1980, bland1987, nicholson1992, quillen1993, sparke1996}.

In section\ \ref{'s.summary'} we provide an overview of observations
of Centaurus~A and relevant
theoretical studies of buoyant bubbles in extragalactic contexts.
Section~\ref{'s.method'} describes the
method of our simulations.
Details of the numerical results and the implications for the physical system
are described in \S\ref{'s.results'}.
A discussion of the main features of this work is given in
\S\ref{'s.discussion'}.

\section{SUMMARY OF RELEVANT OBSERVATIONS \& THEORY}
\label{'s.summary'}

\subsection{Previous observations}

\subsubsection{Galaxy, atmosphere, shells, \& HI clouds}
\label{'ss.galaxy'}

The giant elliptical galaxy Centaurus~A (NGC~5128) is the closest
large radio galaxy, at a distance of $3.6\pm0.2\Mpc$ \citep{soria}
where $1'$ corresponds to a projected distance of $1.0\,\kpc$.
A warped dust and gas disk through the middle of the galaxy
\citep[see \eg ][]{tubbs1980, bland1987, nicholson1992, quillen1992, quillen1993}
is probably a relic of a disk galaxy consumed in a merger.
Stellar shells \citep{malin1983} and HI clouds are distributed
out to distances of several tens of $\kpc$ from the nucleus.
They may also be merger relics.
The cloud locations may be associated with some of the stellar shells
\citep{schiminovich}.
\citet{richter} found the mass of HI clouds around Centaurus~A to be
$7.8\times10^8\times(1.0\pm0.3)\msun$.
\citet{schiminovich} found an HI mass of $4.5\times10^8\msun$ in the disk,
and $1.5\times10^8\msun$ in the shells,
peaking at a projected distance $15\,\kpc$ from the nucleus.

The galaxy has an atmosphere of X-ray emitting gas \citep{feigelson1981}.
\citet{turner1997} observed diffuse X-ray emission within $6'$ of the nucleus,
and concluded that there are two thermal components,
with $kT=0.29\kev$ and $5\kev$.
They suggest that the hotter component may be related to binary stellar systems,
but the possibility that it is thermal gas cannot be excluded
\citep[\eg][]{wagner1996}.
If interpreted as thermal gas,
the sound speeds corresponding to these temperatures are
$2.7\times10^2\,\kms$ and $1.1\times10^3\,\kms$.
The stellar velocity dispersion ranges from
$\sigma_*\approx143\,\kms$ to $90\,\kms$
at locations $2.8\,\kpc$ to $20\,\kpc$ from the nucleus \citep{hui1995}.
This corresponds to a parameter
$\beta\equiv T_*/T=\mu m_p\sigma_*^2/kT = 0.46$
for gas at a temperature of $0.29\kev$ but $\beta=0.03$ if $kT=5\kev$.
The former is in better agreement with the X-ray profile of \citet{forman1985};
it suggests that the stellar velocity dispersion is a modest fraction 
of the virial value.

\citet{marconi2000b} observed the nuclear region in the
near-infrared, where obscuration by the dusty disk is minimal,
and determined the radial profile for the stellar distribution near the core.
Fitting these observations to the empirical ``Nuker law'' model of
\citet{lauer1995},
yielded a core radius of $3''.3 - 5''.6$
with a best fit of $3''.9$, corresponding to $r_0=78\,\pc$.
This should be approximately the value of the core radius of the entire galaxy,
since the stellar component dominates near the nucleus.

The mass of Cen~A was estimated from measurement of the X-ray halo,
to be $6\times10^{11}\msun$ within $<10{\rm kpc}$ by \citet{forman1985}
who assumed a temperature $kT=1.0\kev$.
This should be revised to $1\times10^{11}\msun$ using
the precise temperature of $kT=0.29\kev$ measured by \citet{turner1997}.
Such a value is consistent with the estimate of $1.8\times10^{11}\msun$
within the span of the disk,
and the estimate of $2\times10^{11}\msun$ within $15{\rm kpc}$
by \citet{schiminovich} from velocity measurements of the HI shells,
assuming circular motion.
Kinematic studies of the planetary nebulae \citep{hui1995, mathieu1996}
indicate a mass of $3.1\times10^{11}\msun$ within $25\,\kpc$
and $(3 - 5)\times10^{11}\msun$ within $50\,\kpc$.

\subsubsection{Radio features}

Cen~A possesses radio-emitting features on diverse scales.
A radio jet extends a projected distance
$\sim6\,\kpc$ from the nucleus \citep[\eg][]{jones1996},
and is visible out to $\sim4\,\kpc$ in X-rays
\citep[\eg][]{turner1997}.
Radio plasma from the jets accumulates in inner radio lobes
which extend to projected distances $\sim 8\,\kpc$ from the nucleus,
peak at $\sim6\,\kpc$ and are $\sim4\,\kpc$ across at the widest points.
Vast regions of fainter emission extend beyond the inner lobes.
In projection, the outer lobes comprise the largest object on the sky,
extending across $10\degree$ \citep{cooper}.

\citet{morganti1999} studied the radio emission emanating from the
northern middle lobe and its immediate surroundings.
The NML structure extends to at least $40\,\kpc$ from the nucleus,
and the brightness peaks at $23\,\kpc$ in projection
(see Figure~\ref{'fig.radiologbox'}
for a grey-scale representation of their Figure~2 contour image).

Near the brightest part of the NML \citet{feigelson1981}
detected diffuse soft X-ray emission from a region extending $1\,\kpc$ wide
and a projected height of $5\,\kpc$ in the direction away from the galaxy.
By their interpretation,
if the X-ray emitting region is composed of thermal gas at temperatures
$\sim10^6{\rm K}$,
compressed from the surrounding medium then it must be ephemeral,
with timescales of $\sim10{\rm Myr}$ and $\sim1{\rm Myr}$
for cooling and dynamical expansion respectively.
They also considered an alternative synchrotron model
with an electron population $10^2{\rm Myr}$ old.
\citet{morganti1999} found a $\sim2'$ wide radio-emitting structure
connecting the NML to the inner lobe.
They named this the ``large-scale jet'' and noted that its direction
is aligned within a few degrees of the projected extension of the nuclear jet.

\subsubsection{Optical filaments and star-forming regions}

\citet{blanco1975} discovered several filamentary ionized knots in 
the vicinity of the NML.
Figure~7 of \citet{morganti1999}
shows the positions of the filaments in relation 
to the radio emitting structures
and an HI cloud at the western margin of the NML and ``large-scale jet.''
In projection, the inner filaments start at $7.8\,\kpc$ from the nucleus
(\ie close to the top of the northern lobe)
and extends a further $2\,\kpc$,
in a direction parallel to the ``large-scale jet''.
Similarly, outer filaments exist at $13.6\,\kpc$ in projection,
extending a further $8\,\kpc$ and are coincident with
the southern edge of the NML.
The outer filaments appear projected at the southern edge of the NML,
and on the eastern edge of a $6\times10^6\msun$ HI cloud
\citep{morganti1999}.
Proposed mecahnisms for the excitation of the 
filaments include beamed emission
from the nucleus \citep{morganti1991}
or autoionizing shocks from local disturbances associated with the
radio plasma \citep{sutherland1993}.
The inner filaments show a spread of velocities $\sim360\,\kms$,
up to $\sim430\,\kms$ blue-shifted relative to the galaxy
\citep{morganti1991}, whereas the components of the outer filaments 
are blue-shifted by $\la 300\,\kms$,
with two outliers blue-shifted by $\sim500\,\kms$,
\citep{graham1998}.

\citet{graham1998} observed a concentration of young stars between 
the outer filaments and the HI cloud
near the southwest edge of the NML.
\citet{mould} determined that the stars are of age $\approx16\Myr$.

Both \citet{graham1998} and
\citet{mould} suggest that the star-formation is shock-induced by 
disturbances associated with the NML
and/or the ``large-scale jet''.
The HI cloud has sufficient velocity for it to have traversed
the disturbed region by a distance equivalent to its own width during the 
age of the stars;
it is most likely the source of star-forming gas.

\subsection{Previous simulations of buoyant bubbles}

\citet{churazov} performed hydrodynamic simulations of a rising
buoyant bubble in order to explain the
radio and X-ray morphology of M87 out to a distance of $50\,\kpc$.
They used an Eulerian ZEUS-3D code.
Magnetic fields were assumed to be tangled on small scales,
with a local proportionality between $P$ and $B^2$.
The initial bubble of radio plasma was placed $9\,\kpc$ from the nucleus,
with a radius of $5\,\kpc$.
Its density contrast relative to ambient gas was $\eta=10^{-2}$.
The undisturbed galaxy atmosphere profile was that of \citet{nulsen}.
The $200\times200$ 2D grid extended to a radius of $51\,\kpc$,
in spherical coordinates.
The energy distribution of the radio-emitting relativistic plasma
was determined by using $2000$ tracer particles
to follow the energy loss history within the bubble,
assuming local pressure proportionality
between magnetic field and thermal pressure.

\citet{churazov} found that:
(1) in the absence of surface tension an initially spherical bubble
deforms into a torus as it rises,
on timescales of $\sim10{\rm Myr}$;
(2) thermal gas is drawn up from below the bubble,
and is overdense compared to the ambient gas;
(3) the bubble reaches a maximum height at which its
density equals that of the surrounding medium;
(4) thereupon the bubble spreads laterally, forming a ``pancake'';
(5) convection of the atmosphere surrounding the bubble
flattens the local gradients of density and X-ray emissivity.
The shapes of the simulated mushroom-cloud radio bubble and thermal gas
trunk are qualitatively consistent with the radio and X-ray features
surrounding M87.
(I.e. regions of diffuse X-ray excess
just radially inwards from the radio lobes;
these correspond to thermal gas trunks trailing bubble tori.)

\cite{bruggen} studied a bubble rising through the medium of a galaxy cluster,
using the same method and resolution as
\citet{churazov}
but with different scales, geometric proportions and ambient atmosphere model.
The initial bubble size and location were on the order of
a core radius of the cluster atmosphere.
Results were qualitatively consistent with \citet{churazov},
except that a stalk of radio plasma was left on the axis
behind the ascending bubble.
Elongation of the initial bubble promoted Raleigh-Taylor instabilities
which altered the evolution.
Fluid instabilities that ordinarily distort the bubble shape were inhibited
in simulations including a strong ($B^2\sim P$) magnetic field
that was initially parallel to the bubble surface.

\section{SIMULATION OF THE CENTAURUS~A NORTHERN MIDDLE LOBE}
\label{'s.method'}

\subsection{Method and initial conditions}

Our simulations were performed using a piecewise parabolic method, PPM,
\citep{colella1984}.
The code, PPMLR, is adapted from the VH-1 hydrocode
({\tt http://wonka.physics.ncsu.edu/pub/VH-1/}),
with efficiency improved by \citet{sutherland2001}.
It is an explicitly conservative second-order Gudonov scheme
with Lagrangian advection followed by remapping to a fixed Eulerian grid.

Our basic calculations were carried out on an axially symmetric grid
with a resolution $600\times300$ in the $z$ and $R$ directions respectively.
The physical span of the grid was $1000r_0\times500r_0$.
To test the robustness of results, supplementary calculations were performed
at half the basic resolution.
Further simulations were done at triple the standard resolution
but with a third the standard duration to
investigate some fine morphological details.
Initially the gas density and pressure were distributed according to
an isothermal $\beta$-model with the number density, $n$ described by:
\begin{equation}
n = n_0 \, \left[ 1 + \frac {r^2}{r_0^2} \right]^{-\beta}
\label{e:density}
\end{equation}
Note that the so-called $\beta$-model widely used in the literature is wrong
(see \citet{bicknell1995a,killeen1988a} for a discussion)
so that the value of $\beta=0.75$ used by us
is equal to the \citet{forman1985} value ($\approx 0.5\pm 0.1$) times $3/2$.
The value of $\beta$ implied by the stellar and gas temperatures is
$\beta_g = \mu m_p \sigma_*^2 / kT \approx 0.46$
for $\sigma = 143 \> \kms$ (at approximately 36 core radii)
and $kT \approx 0.29 \> \kev$.
(The stellar population may be at a temperature about $\sim0.6$ times 
the virial temperature.)

The nucleus of the galaxy is the coordinate origin,
positioned halfway along the grid's $z$-axis.
Following equation~(\ref{e:density}), above,
the adopted density and pressure profiles of the interstellar medium (ISM) are:
\begin{equation}
\rho=\rho_0(1+\xi^2)^{-\beta}\ ,
\label{'eq.beta.density'}
\end{equation}
\begin{equation} P=P_0(1+\xi^2)^{-\beta}\ ,
\label{'eq.beta.pressure'}
\end{equation}
where
\begin{equation}
\xi\equiv{{r}\over{r_0}}
\end{equation} is the scaled distance from the nucleus,
and $\rho_0$, $P_0$ are density and pressure values at the nucleus.

We establish initial conditions such that there is initially no flow.
The gravitational and pressure forces balance everywhere.
The gravitational field is thereby inferred from
the $\beta$-model pressure gradient.
That is, the gravitational force per unit mass implied by this model is
\begin{equation} f= {1\over\rho}{{dP}\over{dr}} =
{{-2\beta\xi}\over{1+\xi^2}} {{P_0}\over{\rho_0 r_0}}
\ .
\end{equation}

The bubble is initially located on the $z$-axis at $z_b=60 \, r_0$
with a spherical radius $r_b=20 \, r_0$.
The ratio of the bubble's radius to its height,
$r_b / z_b$ is approximately the same as
the projected proportions of the northern inner lobe of Cen~A,
$r / z \approx 2\,\kpc / 6\,\kpc$.
The pressure and density of the $\beta$-model at the level $\xi_b=z_b/r_0$
are used to set the constant pressure and density everywhere within the bubble.
The bubble pressure and density are prescribed by
\begin{eqnarray} P&=&P_0(1+\xi_b^2)^{-\beta}\ \mbox{and}
\\
\rho&=&\eta\rho_0(1+\xi_b^2)^{-\beta}\ ,
\label{'eq.bubble.density'}
\end{eqnarray}
where $\eta$ is the ratio of the bubble density to the ambient density
at $z=z_b$.
We have investigated cases of bubbles with $\eta=0.5$, $10^{-2}$ and $10^{-3}$.
Figure~\ref{'fig.vfield000'} shows an initial density image of the
galaxy atmosphere and the bubble.

In order to track the different gases
(\ie radio emitting plasma and the ISM),
we have implemented a scalar tracer variable $\varphi$,
which is passively advected with the fluid.
The tracer variable, $\varphi$, is constant along a stream line so that
\begin{equation} {{d\varphi}\over{dt}}\equiv
{{\partial\varphi}\over{\partial t}} +
v_\alpha{{\partial\varphi}\over{\partial x_\alpha}} =0  \ .
\end{equation}
This is implemented in the code in the conservative form:
\begin{equation}
{{\partial}\over{\partial t}}(\rho\varphi)
+{{\partial}\over{\partial x_\alpha}}(\rho\varphi v_\alpha) = 0\ .
\end{equation}
Bubble material initially has a tracer value $\varphi=1$,
and non-bubble gas has $\varphi=0$.
In each cell of the evolving simulation, the value of $\varphi$
therefore indicates
the volumetric fraction of material originating within the initial bubble,
\ie $\varphi$ traces the local distribution of radio plasma
as distinct from thermal gas.

\subsection{Scaling}
\label{'ss.scaling'}

The gas dynamic equations are scaled according to the following prescription.
First, the unit of length is taken as the core radius of the ISM
profile, $r_0$.
However, it is not crucial that we match this exactly
with the observed core radius of the galaxy for the following reason:
The bubble is located well outside a core radius where
the number density and pressure follow a power law in $r/r_0$.
Hence, $r_0$ is in reality a scaling radius
and as we show below (\S\ref{'ss.rescaling'}),
it is necessary to adopt a value of $r_0$
approximately twice the optical core radius
in order to match the simulations and the observed physical scale of the NML.

Second, the velocity scale is the isothermal sound speed of the ISM,
\begin{equation}
v_0=\sqrt{{kT}\over{\mu m_p}}= 210 \, \kms
\left( \frac {kT}{0.29 \, \kev} \right) \ .
\end{equation}

Third, the time scale is
\begin{equation}
t_0\equiv{{r_0}\over{v_0}}=0.25 \, \Myr
\left({{r_0}\over{100\,\pc}}\right)
\left({{kT}\over{\kev}}\right)^{-1/2}\ ,
\label{'eq.time.unit'}
\end{equation}

\section{RESULTS}
\label{'s.results'}

\subsection{Evolution}

\subsubsection{Physical components \& velocity field}
\label{'ss.components'}
\label{'ss.velocities'}

Let us now describe the general characteristics of our simulations.
This description is aided by
the images of density and velocity vector fields in
Figures~\ref{'fig.vfield000'}-\ref{'fig.vfield600'};
and the plots in Figures~\ref{'fig.vmax.eta'}-\ref{'fig.xivbeta'}.

Bubble plasma rises buoyantly in the ambient gas
(density $\rho_{\rm a}$)
with acceleration $\sim(1-\rho_{\rm b}/\rho_{\rm a})f$
($f=$ local gravitational acceleration).
The thermal gas circulates around the bubble
and fills the space behind and beneath the bubble.

As decribed by \citet{churazov}, in the early stages of the simulation
a trunk of dense thermal gas moves up along the $z$-axis
from regions immediately beneath the bubble.
This uplift is driven by a rarefaction
that develops behind the rising bubble and which persists, albeit weaker,
as the bubble develops into a torus.
It penetrates the bubble from below,
and makes contact with the thermal gas on the upper side,
(\eg Figures~\ref{'fig.vfield060'}, \ref{'fig.vfield150'}),
thereby turning the bubble from a sphere to the shape of a mushroom-cap
and then into a torus.
When the trunk breaks through the top of the bubble
(\eg Figure~\ref{'fig.vfield060'}),
velocities at the interface between the uplifted thermal gas
and the bubble can temporarily exceed the sound speed of the ambient gas,
as indicated by the early peak in $|v_{\rm max}|$
in each of the curves of maximum velocity against time in
Figure~\ref{'fig.vmax.eta'}.

The torus rises buoyantly, and the majority of the radio plasma
remains within it throughout our
simulations (to $t=1200t_0$) although there is some peripheral mixing
between the radio plasma and the ambient gas.
At some stages the primary torus has a circular cross-section
(\eg Figure\ \ref{'fig.vfield300'}),
but at other times the cross-section is highly flattened
(\eg Figure\ \ref{'fig.vfield600'}).

As it rises, the torus intermittently sheds minor vortex rings.
These rings occur closer to the galaxy,
have lower density contrasts relative to their surroundings,
and typically propagate down towards the nucleus of the galaxy.

Thermal gas circulates around the main bubble torus and also around
many of the blobs of plasma that separate from the main bubble.
The disturbed gas is drawn towards the $z$-axis
on the underside of the blob,
vertically upwards on the inner radial side,
and radially outwards across the upper surface.
The velocity of the vortex ring increases towards the nearest torus surface,
but drops to zero in the central eye of the circulation
(see \eg the prominent vortex rings in Figure~\ref{'fig.vfield300'}).
At late stages ($t\ga200t_0$)
the maximum velocities are typically $\sim0.7v_0$
(Figure~\ref{'fig.vmax.eta'}),
which is a large fraction of the adiabatic sound speed, $\sqrt{ 5/3} v_0$.
Greater maximum velocities  occur at earlier times
in the cases of $\eta=10^{-2}, 10^{-3}$;
the maximum velocities are typically $\approx1.2v_0$ until $t\approx300t_0$
(\eg the system in Figure~\ref{'fig.vfield060'}
where the thermal gas trunk has just broken through the bubble).

The early-stage thermal gas trunk persists as
a local region of overdense, rapidly uplifted gas
immediately beneath and bounded by the major torus.
At times $t\sim120t_0 - 150t_0$,
soon after the trunk penetrates the top of the mushroom,
the most overdense area of the trunk is typically twice as dense
as the $\beta$-model density at the same $\xi$ coordinate.
In the $\beta=0.75$ model this is
equivalent to the undisturbed density contour at a location
$\approx0.6$ times as close to the nucleus.
(See the clear overdensity of thermal gas in regions
encompassed by the torus in, for example,
Figures~\ref{'fig.vfield150'},\ref{'fig.vfield300'}.)

At sufficiently late stages ($t\ga150t_0$) the torus is high and wide enough
that its uplifting influence on the lower parts of the trunk is weaker.
Beneath the uplift region there is a stagnant region,
and further beneath this
the unsupported gas slumps towards the galaxy under gravity.
The slump is transonic
with approximately the same maximum velocity as the uplift region.
The slumping region is $\ga 40 \, r_0$ away from the major torus.
(See Figure~\ref{'fig.vfield150'}
where a weak, low-velocity slump is commencing close to the galaxy,
and
Figures~\ref{'fig.vfield300'},\ref{'fig.vfield600'}
in which the slump is fully developed
and transonic throughout much of the thermal gas trunk.)

\subsubsection{  Ascent of the radio plasma to the neutral buoyancy point}

\label{'ss:ascent'}

The evolving bubbles rise until reaching a neutral buoyancy point,
where the bubble density equals the ambient gas density.
By equating the undisturbed ISM density (\ref{'eq.beta.density'})
with the initial bubble density (\ref{'eq.bubble.density'}),
a theoretical upper altitude is estimated,
$\xi_{\rm final}=\left[{(1+\xi_b^2)\eta^{-1/\beta}-1}\right]^{1/2}
\sim\xi_b\eta^{-1/2\beta}$.
When $\xi_b=60$ and $\beta=0.75$,
the limit is
$\xi_{\rm final}\approx 95, 1.3\times10^3$ and $6.0\times10^3$
for $\eta=0.5, 0.01$ and $0.001$ respectively.
In practice, these values are upper limits because
mixing and entrainment of thermal gas
may increase the effective value of $\eta$ as radio plasma rises.

Figure\ \ref{'fig.xiveta'} shows the inner ($\xi_{\rm min}$)
and outer ($\xi_{\rm max}$) radial extremes
of the bubble plasma as determined by the distribution of the tracer
$\varphi$, for $\beta=0.75$ and $\eta=0.5, 10^{-2}, 10^{-3}$.
In the case $\eta=0.5$ the bubble achieves neutral buoyancy when
$t\approx400t_0$.
At this stage the median bubble position agrees with the prediction
for material originating in the middle of the initial bubble,
$(\xi_{\rm min}+\xi_{\rm max})/2\approx80\la\xi_{\rm final}$.

The cases $\eta=10^{-2}, 10^{-3}$ are not clearly separated in their
$\xi_{\rm min}$ and $\xi_{\rm max}$ evolutions
because both the respective $\xi_{\rm final}$ values
are far beyond the limits of our numerical grid.
However the lateral spreading of bubble material
(Figure~\ref{'fig.rveta'})
is different in the $\eta=10^{-2}, 10^{-3}$ cases:
for $\eta=10^{-3}$
the radio plasma is more widely distributed perpendicular to the $z$-axis
after $t\approx500t_0$
(compare the upper two curves of Figure\ \ref{'fig.rveta'}).
In general the rate of lateral spreading declines after $t\sim600t_0$
(lower curve of Figure~\ref{'fig.rveta'}),
and the spreading is greater for cases with lower $\eta$.

\subsubsection{Effect of $\beta$}

Figure~\ref{'fig.xivbeta'} shows the time-evolution of
$\xi_{\rm min}, \xi_{\rm max}$
for different galaxy $\beta$-models, with $\eta=10^{-2}$.
The outer front, $\xi_{\rm max}$, is not very sensitive to $\beta$;
however the inner radius $\xi_{\rm min}$ and morphology of bubble plasma
are affected by $\beta$.
In early phase of the bubble evolution
$\xi_{\rm min}$ increases with time as the bubble rises
and is penetrated by the thermal trunk.
For $\beta=0.75, 0.9$ a droplet of radio plasma remains on the
axis in the middle of the thermal trunk,
and it sinks slowly until the time $t\approx250t_0$
when it is too dispersed to be resolved numerically in $\varphi$ data.
No such droplet appears in results for lower $\beta$ values.

\label{'def.stalk'}
At some medium time ($t>200t_0$) in the evolution of each bubble,
the torus has widened and a wisp of radio plasma
is far enough from the uplift region
to become caught in the slump of thermal gas on the $z$-axis.
This wisp becomes a rapidly descending radio stalk,
with $r_{\rm min}$ decreasing in time at a rate of $\approx0.5v_0$.
The time of stalk formation is sensitive to $\beta$:
it occurs at $t\approx 210t_0, 250t_0, 400t_0, 600t_0$
for cases of $\beta=0.9, 0.75, 0.5, 0.375$ respectively and $\eta=10^{-2}$.
The stalk descends to a distance from the core
approximately equivalent to the lower surface of the initial bubble.

\subsubsection {Rise and scaling of the bubble}
\label{'ss.rescaling'}

We can compare the distribution of bubble material in simulations
with the distribution of the NML and northern inner lobe,
in order to limit and deduce the plausible ranges of system parameters
and the age of the NML.
With a scale of $r_0=0.10\,\kpc$,
none of our simulations ends with a substantial fraction of the bubble material
at or above the height of the peak radio brightness of the NML,
\ie $23\,\kpc$ from the nucleus in projection
(Figures~\ref{'fig.xiveta'},~\ref{'fig.xivbeta'}).
The bubble with $\eta=0.5$ rises by less than its own initial diameter;
therefore the NML must have a smaller value of $\eta$.
With $\eta=10^{-2}, 10^{-3}$ a factor $\sim3$ rise in the top of the bubble
matter is possible by a time $\sim10^3t_0$,
and a factor $\sim2$ is possible by the time $200t_0$.

In order to match the simulated and observed radio morphologies
(as shown below in \S\ref{'ss.fuzzy.hat'}),
and to have circulation velocities high enough to explain
the velocity spread of the outer filaments (recall \S\ref{'ss.velocities'}),
we propose an age of $<200t_0$.
For a simulated bubble of this age to be at a height corresponding to the NML,
we must rescale the linear unit of the simulation to $r_0\approx0.2\,\kpc$.
This implies that the NML precursor
must have been roughly twice as big and far from the galaxy
as the present inner lobe.
Such rescaling is valid because at all stages of the bubble's evolution
it is in the power-law region of the ISM structure,
far from the core (\S\ref{'ss.scaling'}).
The correspondingly revised time unit is $t_0\approx0.92\Myr$.

\subsection{Simulated radio morphology}

\subsubsection{Method and morphological identification}

Assuming that the magnetic field is tangled on small scales
and that its pressure is  proportional to the gas pressure,
the synchrotron volume emissivity per unit frequency is
$j_\nu\propto\varphi P^{(3+\alpha)/2}$,
where $\alpha\approx0.6$ is the spectral index.
Surface brightness maps are rendered by revolving the axially symmetric
pressure and tracer data into three-dimensional cylindrical structures,
and then numerically integrating $j_\nu$ along the line of sight
for each pixel of the surface brightness image.

We define characteristic stages of the simulated evolving radio morphology,
as depicted in high resolution in Figure~\ref{'fig.hires'},
and with comparisons in $\eta$ in Figure~\ref{'fig.eta.morphologies'}.
these are compared with observational radio images \citep{morganti1999}
to indicate approximately how long ago the NML began its buoyant rise.
The evolutionary stages we define can be described as:
\begin{itemize}
\item ``initial bubble'':
limb-darkened circle;
$t=0$
(\eg see top-left panel of Figure~\ref{'fig.hires'});
\item ``mushroom cap'':
limb-darkened mushroom shape, dimpled at the bottom,
as the thermal gas trunk penetrates the side closest to the galaxy
and sweeps radio plasma upwards on the $z$-axis;
the edge furthest from the galaxy points in the positive $z$ direction
with fainter emission;
$t\la80 \, t_0$
(\eg see $t=60t_0$ panel of Figure~\ref{'fig.hires'});
\item ``young torus'':
thermal trunk has completely penetrated the bubble, deforming it into a torus;
bright torus is narrow compared with the central hole;
a wispy `hat' of emission extends above the main torus;
$80t_0\la t \la 200t_0$
(\eg see $t=140 \, t_0$ panel of Figure~\ref{'fig.hires'});
\item ``torus + stalk'':
wider bright torus; some faint skirts close
to the torus, above and mainly
below; with wispy emission from a stalk-like feature descending on
the $z$-axis;
$t \, \sim10^{2.5}t_0$
(\eg see $t=380 \, t_0$ panel of Figure~\ref{'fig.hires'});
\item ``old torus + stalk + rings'':
wider, higher torus, with a stalk and numerous wispy rings of emission
partially filling the space between the main torus and the galaxy;
highest extent of bubble material
is $\approx 3$ times further from galaxy than in the initial condition;
$t\sim10^3 \, t_0$
(\eg see the $t=1200 \, t_0$ panels of Figure~\ref{'fig.eta.morphologies'})
\end{itemize}

\subsubsection{Appearance of the torus and inclination}

At stages with a well-defined torus,
the apparent morphology depends on orientation.
When the $z$-axis is at a small or medium angle to the line of sight
(inclination $\la 70\degree$)
the torus appears as a bright ring
with an aspect ratio depending on inclination.
Thus \citet{churazov} identify rings observed in radio images of M87
with buoyantly rising tori (\eg in their Figure~1)
viewed at an angle $\sim 45^\circ$.
When the torus is viewed almost edge-on (inclination $\ga 70\degree$)
it appears as a bright bar transverse to the $z$-axis,
with darkened edges and a somewhat darkened middle
(see Figure~\ref{'fig.inclination'}).
The brightest part of the NML of Cen~A
appears as a transverse bar in the radio images of \citet{morganti1999}
(see Figure~\ref{'fig.radiologbox'}).
If, as our model suggests, this is a buoyant torus,
then its apparent aspect ratio implies an inclination $\approx73\degree$.
This estimated inclination is within the observational limits of
\citet{tingay1998}
for the direction of the radio jet on parsec scales.
Furthermore, if the NML is at a toroidal stage, its age has a lower limit,
$t \ga 80 \, t_0$.

\subsubsection{Fainter radio features}
\label{'ss.fuzzy.hat'}

Radio-emitting wisps and skirts occur when radio plasma is sheared
from the torus and entrained into the thermal gas.
Wisps complicate the morphology increasingly in later stages.
Firstly, as the mushroom cap is deformed into a ``young torus,''
some radio plasma displaced by the upthrusting thermal trunk
is redistributed above the torus,
resembling a ``fuzzy hat'' of faint emission above the bar
(see \eg the case of $t=150 \> t_0$, $\eta=10^{-3}$ in
Figure~\ref{'fig.eta.morphologies'}).
As gas circulates around the widening torus,
wisps are drawn outwards across the top of the torus,
downwards across the outside,
and inwards towards the $z$-axis along the underside of the torus.
At some times the wisps are
evenly distributed above and below the torus
(see \eg the cases of $t=300 \, t_0$ in Figure~\ref{'fig.eta.morphologies'}).
Ultimately, in the ``old torus'' stage,
the space between the torus and the galaxy
becomes crowded with radio-emitting vortex rings and skirts
that descended after being sheared off the torus
(see \eg the cases of $t=1200t_0$ in Figure~\ref{'fig.eta.morphologies'}).

In detail the occurrence of minor radio-emitting wisps is sensitive to $\eta$,
the initial geometry and ambient gas/potential profile.
For instance the cases of $\eta=10^{-2}$ and $\eta=10^{-3}$
but the same initial geometry
show different particular distributions of wispy skirts by the late stages,
as seen in the last columns of the central and lower rows of
Figure~\ref{'fig.eta.morphologies'}.
\citet{churazov} recognised this sensitivity,
which accounts for the large skirt or secondary torus
seen in their simulation,
which was more prominent than any of minor rings that we obtained.

The \citet{morganti1999} radio images of the NML show an extended
region of faint emission  above the bar (torus).
This morphology most closely resembles that of the ``young torus''
with its ``fuzzy hat.''
The height of the faint emission region is greater
in comparison with the bar than we have found in our simulations
(typically the simulated ``fuzzy hat'' reaches a maximum height that is
approximately the width of the torus).
The excess may be due to details of the initial geometry;
a prolate rather than spherical initial bubble
might produce a taller ``hat.''
It may also be affected by the relative pressure
between the intial bubble and the ISM.
We note that the simulated radio morphologies
at earlier and later stages lack extended diffuse emission above the torus.
Although the comparison between the observed and simulated morphologies
is not perfect, the simulation strongly suggests that
the dynamical age of the NML is
$t\approx 120 - 150 \, t_0$.

\subsubsection{A radio stalk and its motion}

The descending body of radio plasma mentioned in \S\ref{'def.stalk'} 
at late times is a stalk on the $z$-axis.
Its thickness is a few $r_0$, which is a fraction of the 
thickness of the column of slumping thermal gas carrying it.
No such stalk appeared in the simulations of the M87 radio lobes
by \citet{churazov},
which likely indicates that such fine features are 
sensitive to geometry and system parameters.
\citet{bruggen} found a radio stalk in simulations of bubbles on 
galaxy cluster scales,
however those stalks appear when part of the bubble contracts laterally
about its initial position.
The stalks in our simulations involve radio plasma
moving into regions previously devoid of bubble material.

In medium and late times with $\eta=10^{-2}, 10^{-3}$
(\eg $t=600t_0, 1200t_0$ in Figure~\ref{'fig.eta.morphologies'})
transverse rings with diameters 
of $\sim 10 \, r_0$ propagate vertically up and down the stalk.
It is not clear whether these features
represent vertical transport of radio plasma,
or a vertically propagating wave in the stalk thickness.
In an off-perpendicular view the
rings overlap, giving the appearance of a fat stalk with bright core 
and faint margins.
We speculate that three-dimensional asymmetries of the rings might give 
qualitatively similar images at all orientations.
This fat-stalk morphology has thickness comparable to 
the ``large-scale jet'' found by
\citet{morganti1999}.

However, the simulated radio stalk cannot be a complete description 
of the ``large-scale jet.'' 
The morphology of the NML alone is that of a ``young torus'' 
($t\la150t_0$), which occurs before slumping
commences in the thermal gas trunk.
At this age, only uplift occurs.

The observed velocities of both the inner and outer filaments,
which are respectively near the inner lobe 
and NML, are blue-shifted
\citep{graham1998, morganti1991},
indicating upflow in the regions at both ends of the ``large-scale jet''.
We speculate that radio plasma from the inner lobe is driven 
by the same pressure gradient
that causes the upflow, thereby achieving a similar momentum flux.

Its velocity will be higher because
it is lighter with the velocity scaling approximately as the square 
root of the density ratio of thermal gas to radio plasma.
The diameter of the uplifted region is 
determined by the extent of a low pressure
region that is formed at the middle of the rising torus.
Hence the diameter of the ``large-scale jet''
($3~kpc$) and the thermal trunk should be similar.
At $t=150 \, t_0$ the width of the thermal trunk is 5.0~kpc,
based upon the point where the density becomes equal to the background density.
Thus the width of the trunk in the simulation and the observed
width of the ``large-scale jet'' are similar.

Another possibility is that the jet feeding the NML is not cut off 
completely and that the large-scale
jet represents a reduced historical level of jet activity.

\subsection{Physical implications}

\subsubsection{X-ray excess coinciding with the NML}
\label{'ss.xray.excess'}

In the ``young torus'' stage of our simulations, the overdense trunk 
of thermal gas bounded by and
immediately beneath the radio plasma torus is up to a factor $2$ 
times as dense as the undisturbed gas
at equivalent locations in the potential (\S\ref{'ss.galaxy'}).
As the X-ray emissivity is proportional to
the square of the density, the overdense trunk produces a local 
excess of X-ray emission.

We identify this feature with the elongated, diffuse X-ray emission 
coincident with the NML from $15'$
to over $20'$ from the nucleus, discovered by \citet{feigelson1981}.

We correct their estimate of the
region's luminosity in the $0.5-4.5\kev$ band,
$L_{0.5-4.5\kev}=1\times10^{39}{\rm erg} {\rm s}^{-1}$ for a distance 
of $3.5\Mpc$ rather than the $5\Mpc$ they assumed.
The region has approximately the volume of a 
cylinder $1'$ in diameter and $5'$ high.
Using the MAPPINGS~III code (v1.0.0m) we calculated the 
$0.5-4.5\kev$ X-ray emissivity at a temperature $0.29\kev$,
$2.1 \times10^{-23}(n_{\rm H}/\pcm3)^2\ {\rm erg} {\rm s}^{-1}\pcm3$, 
from which we infer the number density of the thermal gas to be
$n_{\rm H}\sim1.4 \times 10^{-2} \pcm3$.
By extrapolating the radial density profile of
\citet{forman1985}, and using densities near knots of the jet derived 
by \citet{schreier1981}, with a correction to a temperature $kT=0.29\kev$,
we find that undisturbed ISM should have a density of
$\sim 1\times10^{-3}\pcm3$
at positions equivalent to the outer filaments, 
or $\sim 2\times10^{-3}\pcm3$ at the inner filaments.
Given the above density, the cooling time of the 
thermal trunk is $\approx 50 \> \Myr$.
Thus the gas will not cool significantly on the timescale
$\approx 140 \Myr$ that we estimate below (\S\ref{'ss.sfr'}).

\citet{churazov} noted the connection between simulated thermal gas 
trunks and the elongated regions of
X-ray excess trailing the radio lobes of M87.
Those X-ray features appear longer in proportion to the
radio lobes (lower-left panel of Figure~1 in \citet{churazov}) than 
is the case for Cen~A (Figure~6 of \citet{feigelson1981}).
This would be consistent with the M87 radio 
lobes being dynamically older than the NML of Cen~A:
at times of a few hundred $t_0$ the thermal gas 
trunk is proportionately longer,
and the hole in the radio torus more pronounced
(\eg see our Figure~\ref{'fig.vfield300'})
than at the best morphological match for Cen~A
($t\approx150t_0$, Figure~\ref{'fig.vfield150'});
there is a lack of diffuse ``fuzzy hat'' radio emission above the torus.
Adopting the approximate temperature of the ISM in M87
and initial bubble position of
\citet{churazov}, the time scale of M87 is $t_0\approx0.38\Myr$.
The preferred age of the M87 torus,
$t\sim8\times10^7\yr$, corresponds to $\ga210t_0$, which indeed is 
consistent with the M87 lobe being
more dynamically evolved than that of Cen~A.

Our simulations yield an overdensity of $\sim 2$ times in the 
uplifted thermal gas trunk,
relative to the undisturbed X-ray emitting atmosphere at an equivalent distance 
from the nucleus of the galaxy.
However the ratio of the trunk to ambient densities, as calculated 
above from the observations of
\citet{feigelson1981, forman1985, schreier1981}, is $\sim 7$.
We note however, that in \citet{feigelson1981}
the data were not presented in such a way as 
to facilitate an independent estimate of the trunk density;
the density of the background medium is an extrapolation.
Further, high resolution {\em Chandra} observations
are required for a definitive assessment.
On the other hand,
it is possible that bubbles that are initially larger may raise a trunk 
more effectively,
resulting in a greater overdensity factor.
We have partially confirmed in a simulation with a bubble
that is larger by a factor of 1.5.
This results in a maximum overdensity of 3.

\subsubsection{A nonspherical gravitational potential}

The radio structure of the NML (Figure~\ref{'fig.radiologbox'})
observed by \citet{morganti1999} bends towards the north.
Whereas the ``large-scale jet'' points
approximately outwards away from the nucleus,
the inferred torus is tilted by several degrees relative to the
central direction and the tapering diffuse emission above it is tilted further,
so that it almost points due north.
We interpret this bend
as evidence for a non-spherical gravitational potential.
The equipotential surfaces become more prolate
or oblate with distance from the nucleus,
thereby providing a buoyant force that acts in different directions
at different locations.
This is an independent probe of the gravitational potential at large
distances from the visible galaxy.

\subsubsection{The outer filaments}
\label{'ss.filaments'}

In this and the following subsections we present a model for the
formation of the outer filamaents and
the nearby star-formation  region that utilises the velocity field in
the X-ray emitting thermal gas caused by the rising torus.

The large HI cloud near the NML and outer filaments
\citep{schiminovich} has a peak HI column density of
$\sim10^{21}\pcm2$ and typical width $\sim3\,\kpc$,
implying an average number density $n_{\rm H} \sim 0.1\> \pcm3$.
Local regions may have densities greater or lesser than the average.
The HI cloud has a blueshift of $\sim200\,\kms$ with respect to the 
systemic velocity of the galaxy
\citet{schiminovich}.

In projection, the eastern edge of the cloud is adjacent to the
``large-scale jet'' \citep{morganti1999},
the region of young stars and the outer filaments \citep{graham1998, mould}.
Thus, the large cloud
appears to have entered the region affected by the overdense region
affected by the rising torus of the NML.
Gas on the disturbed edge of the cloud is shock-compressed to
form stars (\S\ref{'ss.sfr'}) and the
flow also ablates gas from the edge.
These initially cold and dense
clouds are accelerated and shredded
by the Kelvin-Helmholtz and Rayleigh-Taylor instabilities in a
cascading process that results in a
stream of small cloudlets that are eventually accelerated to an
appreciable fraction of the free-stream velocity.

The ablated clouds initially move
at the orbital velocity of the large HI cloud,
(i.e. are blusehifted with a velocity $\sim 200 \> \kms$)
approximately perpendicular to the X-ray gas upflow.
The ram-pressure acceleration of clouds by the upflow
increases the velocity dispersion.
This is consistent with the velocity dispersion of order the
atmospheric sound speed \citep{graham1998}.
For a spherical cloud with a density of $\rho_c$ and a radius of $R_c$
driven by a medium with a density of $\rho_a$ moving at a velocity $V_a$,
the drag force is
$F_D=C_D\pi R_c^2\rho_a V^2
= (4\pi/3)R_c^3\rho_c (dv_c/dt)$
where $C_D\sim1$ is a drag coefficient.
Thus the acceleration timescale, $t_{\rm acc}\approx V(dv_c/dt)^{-1}$
is approximately
\begin{equation} t_{\rm acc}
\approx5.4\times10^4\yr {{\rho_{\rm c}}\over{\rho_{\rm a} C_D}}
\left({{R_c}\over{20\,\pc}}\right)
\left({{V_a}\over{270\,\kms}}\right)^{-1}
\ .
\end{equation}
The streaming X-ray emitting gas has a density $n_a\sim10^{-2}\pcm3$
\citep{feigelson1981} (see \S\ref{'ss.xray.excess'}).
In observations of the filaments,
\citet{morganti1991} measured the $\sii$ doublet
to be in the low-density limit, implying $n_c<10^2\,\pcm3$,
which is not diagnostic.
The density of the large HI cloud, as the source of the small clouds,
is likely to be give a more realistic value of the cloud density,
$n_c\sim0.1\,\pcm3$.

The total pressure (static plus ram) of the stream
drives a shock into each cloud.
(The HI cloud may be underpressured with respect to the stream,
so that we incorporate the effect of the static pressure.)
Let $c_I$ be the isothermal sound speed of the stream
and $v_{\rm rel}$ be the relative velocity
between the moving cloud and the stream.
Then, the velocity  of the cloud shock is:
\begin{eqnarray}
v_{\rm sh}&\approx&
\sqrt{
{{2}\over{\gamma+1}}
{{P_{\rm a}+\rho_{\rm a}v_{\rm rel}^2}\over{\rho_{\rm c}}}
}
=
\sqrt{
{{2}\over{\gamma+1}}
{{\rho_{\rm a}}\over{\rho_{\rm c}}} (c_I^2+v_{\rm rel}^2)
}
\nonumber \\
&=& 110\left({ {n_{\rm a}}\over{0.01 \> \pcm3} }\right)^{1/2}
\left({ {n_{\rm c}}\over{0.1 \> \pcm3} }\right)^{-1/2}
{{\sqrt{c_I^2+v_{\rm rel}^2}}\over{400 \> \kms}} \, \kms
\ ,
\label{'eq.cloud.shock'}
\end{eqnarray}
If the cloud has no component of velocity in the direction of the stream
then $v_{\rm rel}\approx (270^2+200^2)^{1/2}\approx 340\,\kms$;
for an ambient temperature $kT\approx0.29\,\kev$, $c_I \approx 210 \> \kms$ and
$(c_I^2+v_{\rm rel}^2)^{1/2}\approx 400\,\kms$.
The shock velocity decreases for larger cloud densities and
increases if there is a cloud velocity component towards the nucleus.

The timescale for shredding of a cloud
is usually shorter than the acceleration timescale.
Let the diameter of a cloud in the flow be
$D_{\rm c}$ and the velocity of the cloud shock be $v_{\rm sh}$,
then the shredding timescale of a shocked cloud in the upflow is
$t_{\rm shred}\sim D_{\rm c}/v_{\rm sh} \approx 2 \times 10^5 \>
(D_{\rm c}/20 \, {\rm pc}) (v_{\rm sh}
/100 \, \kms)^{-1} \> \rm yr$ \citep{klein94}.
As a result of the sheredding,
the acceleration time of the clouds becomes progressively shorter.
If the dense cloudlets have a fractal structure and a distribution of sizes
then the clouds will cover a range of velocities and degrees of shredding.
The outer filaments do indeed show such a range of morphologies
-- from large arcs to elongated smears
distributed in the direction away from the galaxy nucleus
\citep{morganti1991, graham1998}.
We attrribute the former to clouds in the initial stages of acceleration;
in this case the shapes of the emission line arcs
trace the outline of the cloud.
We attribute the smeared features to the shredded remnants
of previously accelerated clouds.
At the available resolution in the images of
\citet{morganti1991, graham1998}, the smallest visible knots are
$\sim1-2''$ wide, corresponding to a radial scale $R_c\sim10-20\,\pc$.

We now consider some of the features in the outer filament region
that enable us to estimate the density and cloud shock velocities.
The largest distinct structures amongst
the outer filaments are arcs with $\sim70\,\pc$ radii of curvature
\citep{morganti1991,graham1998}.
By integrating the $\oiii$ surface brightness contours in
\citet{morganti1991} the luminosity of the prominent arc at
the southern end of the outer filaments is found to be
$L_\oiii\approx(1.3 - 2.6)\times10^{36}{\rm erg}{\rm s}^{-1}$.
The filaments' $\oiii/{\rm H}\beta$ ratios $ \sim 3-5$
\citep{graham1998} imply shock velocities
$\sim 80 \la v_{\rm shock} \la 120\,\kms$.
Assuming spherical cap-like geometry for this arc then
the shock surface area of this arc $\approx 2.5\times10^{41}\> \rm cm^2$.
We used MAPPINGS~III to calculate $\oiii$ emissivities
for shock velocities ranging from $80-120\,\kms$,
to infer a number density $n_{\rm  H}\approx 0.3 - 0.9 \pcm3$ in the cloud.
This number density would be less if the shock were corrugated.
Densities at the lower end of this range are consistent with those
required for shock velocities greater than $80 \> \kms$,
this being the threshold value for \oiii emission.

On the other hand, as we have indicated in (\ref{'ss.xray.excess'}),
the present simulations may not have been successful in reproducing 
the observed over-density of uplifted thermal gas.
We have indicated that this may be related to
the intial size of the NML precursor.
On the basis of the current simulations, therfore,
we cannot exclude the possibility that
the ram pressure of the ``large-scale jet''
may play a role in exciting the outer filaments.
Future X-ray observations are crucial to better
constraining the density of hot gas in the vicinity of the NML.

With densities $n_c\sim0.1\,\pcm3$ and
$n_a\sim1.4\times10^{-2}\pcm3$, the acceleration timescales of clouds
with radius $70\,\pc$ are less than $2\Myr$.
The timescale for disruption of a
$70\,\pc$-radius cloud is $\sim1.5\Myr$.
The similarity of these timescales is consistent with a
distribution of cloud morphologies from large, nearly round blobs
to small, smeared filamentary streaks
\citep[see images in][]{morganti1991,graham1998}.

Thus the spatial and velocity distribution of the outer filaments
appears consistent with the time and
length scales of cloudlets in the HI region being swept into the
upflow of X-ray emitting gas beneath the NML torus.
The pressure of the stream on the faces of the largest
blobs may account for their excitation.

The inner filaments may be undergoing the same processes as the outer filaments.
It is noteworthy that
the velocity spread within each component of the inner filaments is
transonic in the X-ray emitting gas, $\la270\,\kms$.
However there must be additional influences because
the spread of velocities in the overall inner filament region
is approximately $360\,\kms$ \citep{morganti1991},
which exceeds the sound speed of a $kT=0.29\kev$ gas.
The detailed hydrodynamics of this region may be affected by the
``large-scale jet'' and the inner lobe,
and this is beyond the scope of the present paper.

\subsubsection{Star formation, merger history, \& the AGN source}
\label{'ss.sfr'}

With a length scale $r_0\approx0.2\,\kpc$ (in accordance with the
height of the inferred torus in the NML,
see Figures~\ref{'fig.xiveta'},~\ref{'fig.xivbeta'}),
and with an ISM gas temperature of $kT=0.29\kev$,
the morphologically inferred bubble age of $t\ approx 150 t_0$,
corresponds to $1.4 \times 10^2\Myr$,
using the value of $t_0 \equiv r_0/v_0$ given by equation
(\ref{'eq.time.unit'}).
At a time, $t = 150\,t_0$,
the velocity of rise of both the $\eta = 10^{-2}$ and $10^{-3}$
bubbles is $0.27 v_0 \approx 60 \> \kms$
(see figure~\ref{'fig.xiveta'}).
The time of transit past the
$\sim1\,\kpc$ region occupied by the \citet{graham1998} field of the
outer filaments is therefore $\approx 16 \> \Myr$,
exactly the age of the young stars
estimated by \citet{mould}.
This strongly reinforces a dynamical connection between the NML
and the formation of the stars.

In our model, star formation is induced by shocks resulting from the
impact of the X-ray gas on the HI cloud adjacent to the NML
\citep[\eg][]{schiminovich}.
The growth time and length scales of a maximally unstable post-shock region
are estimated in equations~(9) -- (11) in \citet{bicknell2000}
on the basis of the theory of \citet{elmegreen1978}.
For an external gas density,
$n_{\rm H,ext} \sim 10^{-2}\pcm3$, and a cloud density, $n_{\rm H} \> \pcm3$
the shock velocity in the cloud (\ref{'eq.cloud.shock'}) is
$v_{\rm shock}\approx 35\, n_{\rm H}^{-1/2} \> \kms$,
The time scale for maximal growth is
$t_{\rm mgr}\sim 6 n_{\rm H}^{-1/2} \Myr$.
After a time of $t_6 \> \Myr$,
the maximally growing wavelength,
$\lambda_{\rm mgr} \sim 4 n_{\rm H}^{1/2} t_6\,\pc$ is well
within the scale of the star-forming region.
Thus, consistent with the observations of \citet{mould},
significant star formation can occur within a few Myr,
especially if $n_{\rm H} > 0.1$.
This suggests that the emission line region is adjacent to
a star-forming region with somewhat greater density.

The estimate of the age of the NML $\sim1.4\times10^2\Myr$ is significant.
We suggest that this corresponds to
the disruption of radio activity within Cen~A
as a result of the merger that created the present dust lane.
The detachment of the NML precursor from the
presently growing inner lobe implies
that jet activity was absent or greatly reduced during an interval
$\sim 10^2 \> \Myr$.
The duration of this interruption is on the order of
a dynamical time-scale of gas within galaxy.
Moreover this time-scale is typical of the proposed duration of the merger of
a spiral galaxy into NGC~5128,
and the subsequent evolution of the warped dust-lane disk.
\citet{tubbs1980} performed test-particle kinematic studies with an
initially flat disk warping due to
differential precession in a prolate galaxian potential.
The presently observed disk warp was matched
at a simulated age of $2.3\times10^2\Myr$
(after correcting the galaxy's distance to $3.5\Mpc$).
However in further geometric models of initially flat disks,
comprised of mutually-gravitating rings, in an oblate potential,
\citet{sparke1996} determined an age $\sim1{\rm Gyr}$.
\citet{quillen1993} considered the infall and tidal stripping of a
spiral galaxy, starting at a time
when the spiral galaxy was $15'$ from the nucleus of NGC~5128.
They found that a non-flat disk could be
produced directly due to the varying angular momentum of the tidally
stripped material, and that the present disk morphology can occur as early as
$1.6\times10^2\Myr$ after the start of the merger.
This latter scenario is most consistent with our age for the NML.

We speculate that the jet was disrupted by the merger event and that
a large flow of gas to the nucleus
on the merger timescale and either interrupted the accretion flow
into the nucleus or obstructed the jets out of the nucleus.
The time elapsed between the detachment of
the NML precursor and the formation
of the new inner lobe supports the model of \citet{quillen1993}.

\section{DISCUSSION AND CONCLUSIONS}
\label{'s.discussion'}

Following the simulations of buoyant radio-emitting bubbles by
\citet{churazov} and \citet{bruggen},
we have investigated the relevance of such a model
to the Northern Middle Lobe of Centaurus~A,
examining the hypothesis that it originated
as a radio lobe created in an earlier episode of jet activity.
In so doing we have extended these papers in the following ways:
\begin{itemize}
\item
We have used an ISM model appropriate for Cen~A and examined
the importance of the $\beta$ parameter on the dynamics of the bubble.
\item
We have used an initial geometry for the bubble that is typical of
the inner radio lobe of Cen~A,
as a precursor to the present middle lobe, albeit a factor of two larger
in size and distance from the nucleus.
(\citet{churazov} used a bubble over twice as large in
comparison to its displacement from the
nucleus, whereas \citet{bruggen} studied a bubbles starting within a
core radius away from the origin of the potential).
\item
We have examined in detail the effect of the density contrast between
the radio bubble and the ambient medium,
for values of the density contrast, $\eta$, as low as $10^{-3}$.
We found that some aspects of the bubble evolution are
sensitive to the difference between
$\eta=10^{-2}$ and $\eta=10^{-3}$.
For example, lateral spreading of the torus
(Figure~\ref{'fig.rveta'}) and the time at
which the maximum convection velocity occurs (Figure~\ref{'fig.vmax.eta'}).
\end{itemize}

In the light of these simulations, we have proposed a model in which
the precursor to the northern middle lobe of Centaurus~A
grew in an episode of jet activity which ceased some time ago.
At the time when the jet feeding the lobe was interrupted,
the lobe was approximately  twice as large
and twice as distant from the nucleus as the present day inner lobe.
Since then it has been rising buoyantly away from the nucleus.
(with a density contrast $\eta<10^{-2}$ relative to its surroundings).
This model explains the gross features of the NML,
namely its bright base and the fainter radio structure above it.
A more detailed comparison with model and observation would require
fine-tuning of the initial parameters.
For example, one could investigate an elliptical configuration
for the initial structure of the NML
or the effects of an elliptical potential.
Nevertheless, comparison of the gross features of the radio morphology
with the simulations is sufficient for us to date the NML
at $\sim1.4\times10^2\Myr$.
We have suggested that the time of
interruption of jet activity corresponds to
the time at which the  often-discussed merger took place in
Centaurus~A and we regard the
\citet{quillen1993} estimate of $1.6\times10^2\Myr$ for the time it
has taken the NGC~5128 disk to settle into its present configuration
as compelling confirmation of this idea.
This is significant evidence that mergers can disrupt existing AGN
activity, possibly by their effect on the mass transfer to the nucleus.
If a $\kpc$-scale accretion flow is disrupted in this way,
then it takes some time (a few orbital timescales)
for the gas that has been captured by the galaxy
to settle again into a steady accretion flow.
Alternatively, if the jet is smothered by
clouds of material accreted from the consumed disk galaxy
then a similar timescale may elapse
before the matter settles away from the jet axis.
We suggest that this has occurred in Centaurus~A,
leading to the renewed onset of radio activity represented by the 
inner lobe and jet.

This conclusion may be related to other observations of young radio
galaxies and may well provide a
fresh way of looking at the relationship between mergers and radio activity.
For example
\citet{perlman01a} have found evidence for old ($\sim 10^8 \> \yr$)
merging activity in three young
Compact Symmetric Object (CSO) radio galaxies and have suggested that
the mergers may be related to a
delayed onset of radio activity.
The parallels with Centaurus~A are obvious.

The bubble model explains the gross features in radio and X-rays on
the scale of the middle and inner radio lobes.
The rising bubble is surrounded by a transonic
circulation in the denser, ambient, X-ray emitting medium.
The vortex ring pattern is substantially larger than
the extent of radio-emitting plasma and follows the bubble as it rises.
Wisps of radio-emitting plasma are drawn off the bubble
by the circulating gas and extend high above the main part of the bubble
at the present stage.
A trunk of overdense X-ray emitting gas is drawn up from below the bubble,
deforming it into a torus.
The overdense trunk is identified with the diffuse X-ray excess
coinciding with the NML discovered by
\citet{feigelson1981}.
Although the ensuing phase of renewed jet
activity is not addressed by our
simulations we have suggested that the pressure gradient in the
upflow region beneath the NML drew light
radio plasma from the top of the newly forming inner lobe.
This is our suggestion for the ``large-scale jet''
crossing the gap between the inner and middle radio lobes.
Nevertheless, this suggestion remains to be confirmed
by future simulations involving two lobes.

In our model, the evolution of the buoyant radio-plasma bubble on
$\kpc$ scales also accounts for features observed on scales of tens
of parsecs or less.
About $\sim16\Myr$ ago the convection flow caused by the bubble interacted
with a massive HI cloud, resulting in both shock-induced star formation
and ablation of gas from the periphery of the cloud.
Thus we have proposed that the outer filaments of emission line gas
are clouds of a few $\pc$ to $\sim150\,\pc$
thickness swept into the transonic flow and shock-excited by the
dynamic pressure of the stream.
The density and shock velocity estimates are reasonably consistent here
with the proviso that the {\em estimated} density be used.
In this context we have noted the
importance of new X-ray observations of
the NML of Centaurus~A and the need for further simulations with
different geometry.

The velocity dispersion in the ablated clouds results from their
different sizes and different times of ablation.
The inner filaments may also be affected by the thermal upflow.
However, their large velocity spread $\sim 360\,\kms$
is not reproduced by our simulations motivating further work involving the
effect of the initial geometry on the structure of the flow.

\section*{ACKNOWLEDGEMENTS}

We thank R.~Morganti for the radio image of the northern middle lobe
of Centaurus~A.
This work was
supported by an Australian Research Council Large Grant, A69905341



\input{bbl.tex}
\input{figs}

\end{document}

%% file: commands.tex
\newcommand{\GS}{\lower.5ex\hbox{$\buildrel>\over\sim$}}
\newcommand{\LS}{\lower.5ex\hbox{$\buildrel<\over\sim$}}

\newcommand{\eg}{e.g.\ }
\newcommand{\ie}{i.e.\ }

\newcommand{\msun}{M_{_\odot}}
\newcommand{\yr}{{\rm yr}}
\newcommand{\Myr}{{\rm Myr}}
\newcommand{\pc}{{\rm pc}}
\newcommand{\kpc}{{\rm kpc}}
\newcommand{\Mpc}{{\rm Mpc}}
\newcommand{\kms}{{\rm km}\ {\rm s}^{-1}}
\newcommand{\pcm}[1]{{\rm cm}^{-#1}}
\newcommand{\kev}{{\rm keV}}
\newcommand{\degree}{^{\circ}}

\newcommand{\oiii}{{\rm [O{\small III}]}}
\newcommand{\sii}{{\rm [S{\small II}]}}

%% file: figs.tex
\begin{figure}
\begin{center}
$
\begin{array}{c}
\plotone{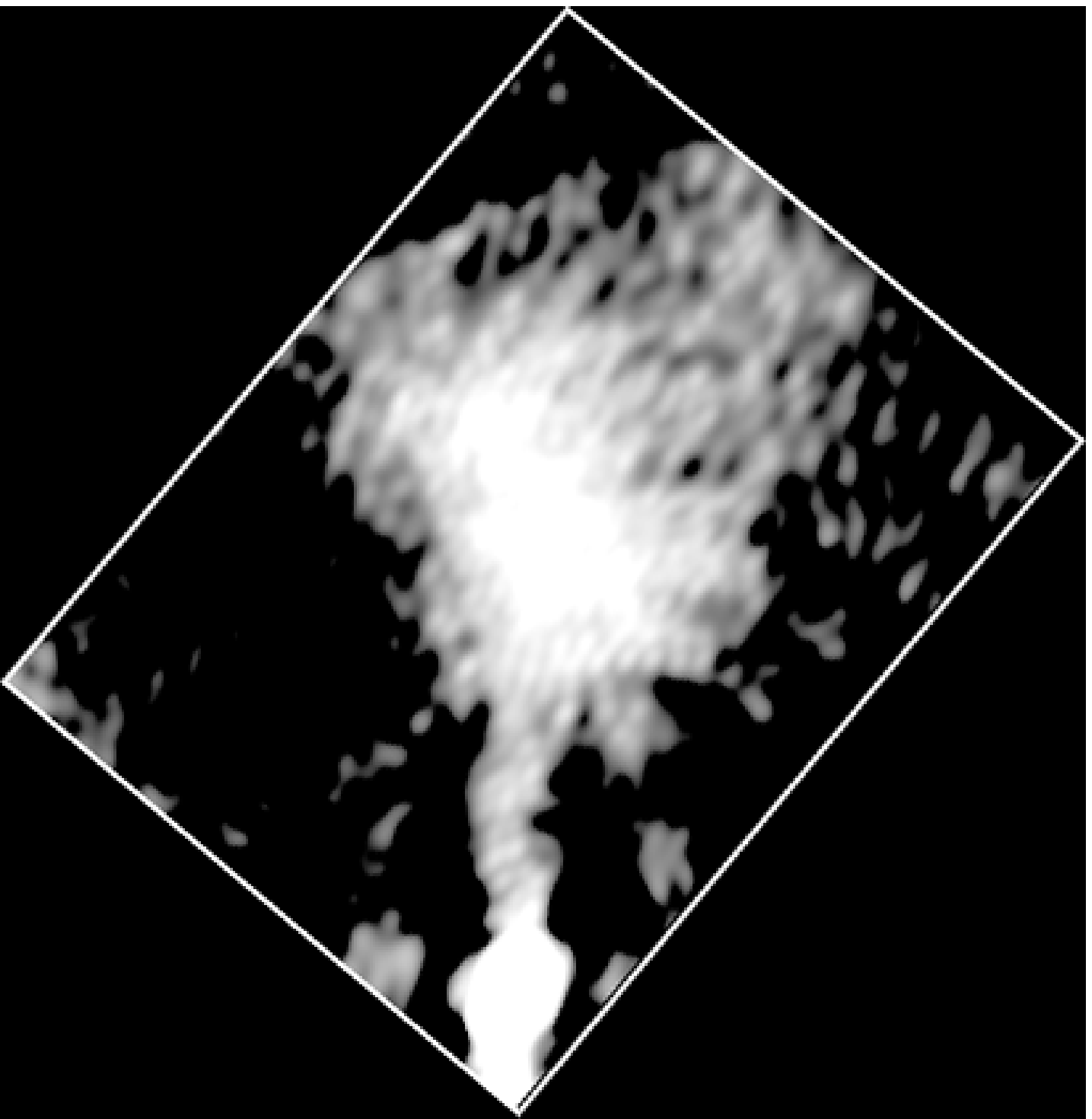}
\end{array}
$
\caption{ Logarithmic radio image of the Cen~A corresponding to
Figure~2 of \citet{morganti1999}.
The inner lobe (bottom) is connected to the NML through a ``large-scale jet.''
Sides of the white box are aligned with the RA and DEC axes;
the image is rotated so that the axis of the ``large-scale jet'' is upwards.
By permission of Dr R.
Morganti.
}
\label{'fig.radiologbox'}
\end{center}
\end{figure}

\begin{figure}
\begin{center}
$
\begin{array}{c}
\plotone{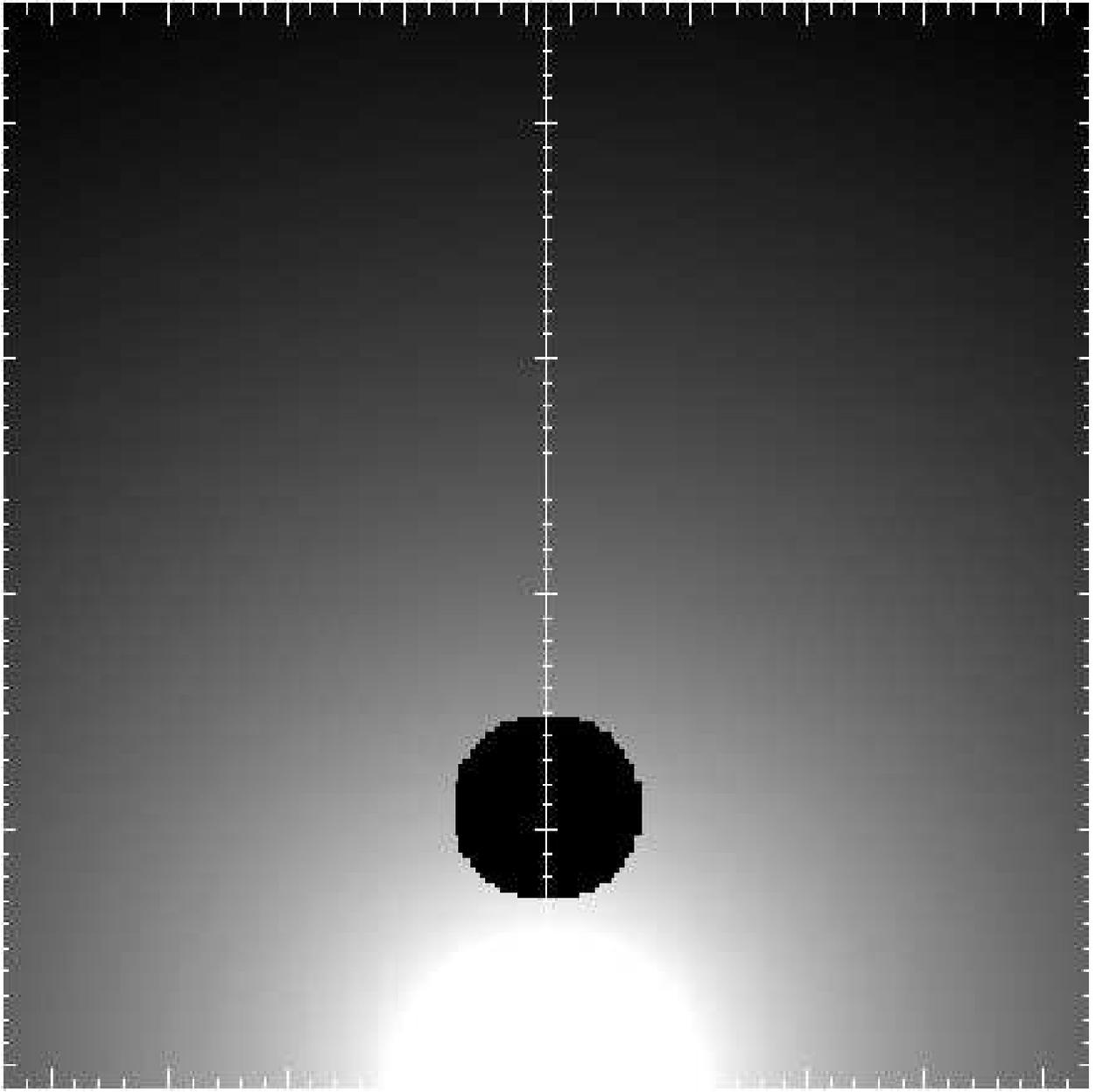}
\end{array}
$
\caption{ Initial density field of the $\eta=10^{-3}$ simulation,
with triple the standard spatial
resolution.
Greyscale brightness is proportional to $\log_{10}\rho$,
with saturation at a fixed level near the nucleus.
The subregion shown here has dimensions
$140\times70$ pixels, (or $\approx
233r_0\times117r_0$) in the cylindrical $z$ and $r$ directions.
The
$z$ axis is directed upwards;
the galaxy nucleus is at the origin,
at the bottom edge of the frame.
}
\label{'fig.vfield000'}
\end{center}
\end{figure}

\begin{figure}
\begin{center}
$
\begin{array}{c}
\plotone{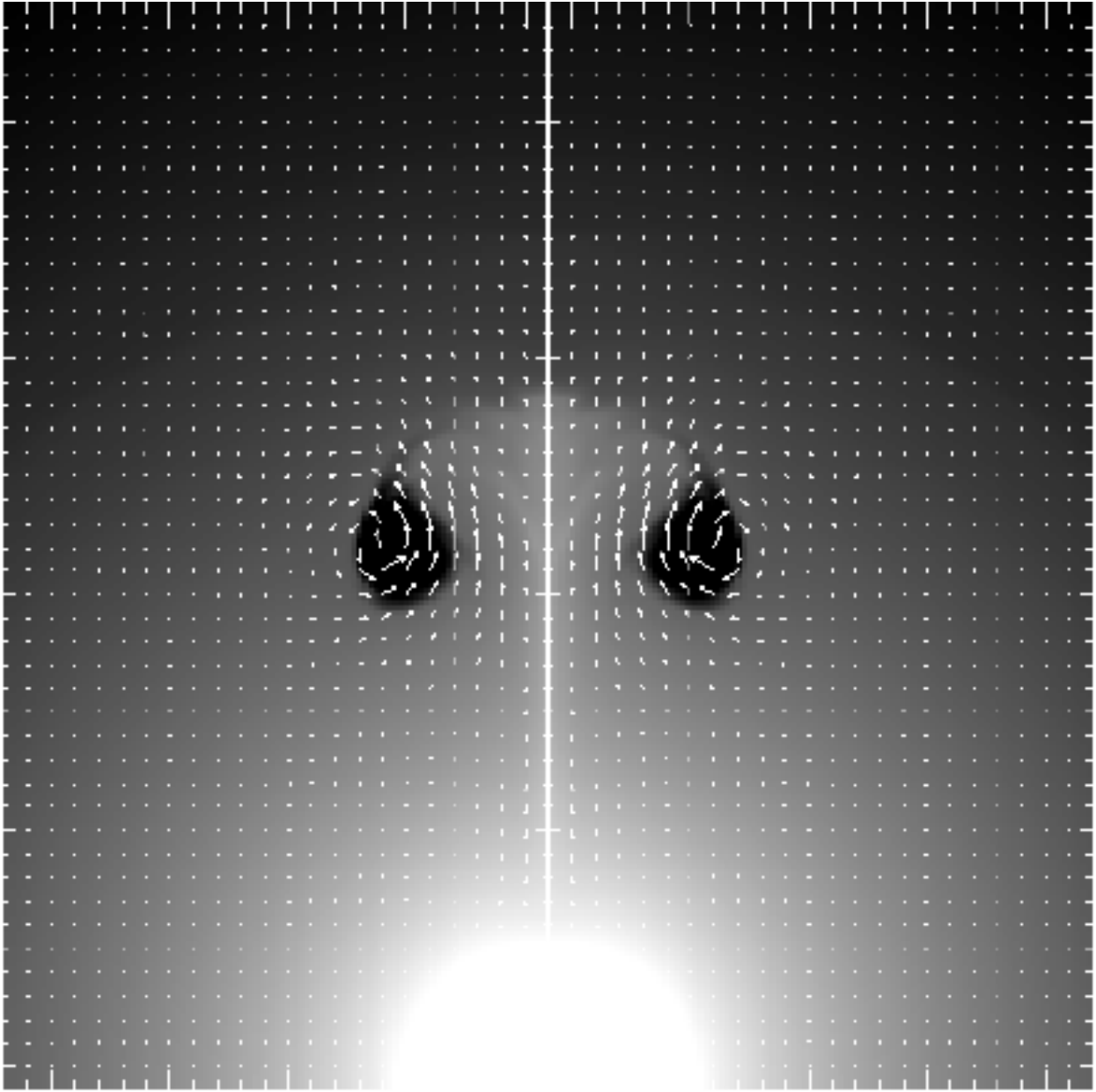}
\end{array}
$
\caption{
Flow velocity vectors at time $t=60t_0$
in the $\eta=10^{-3}$ simulation with triple the standard spatial resolution.
The maximum velocity is $\approx1.2v_0$.
The region and density scale are the same as in Figure~\ref{'fig.vfield000'}.
The thermal gas trunk has almost broken through the initial bubble;
a shock has propagated outwards from the initial shock position and 
appears as a circular density discontinuity.
}
\label{'fig.vfield060'}
\end{center}
\end{figure}

\begin{figure}
\begin{center}
$
\begin{array}{c}
\plotone{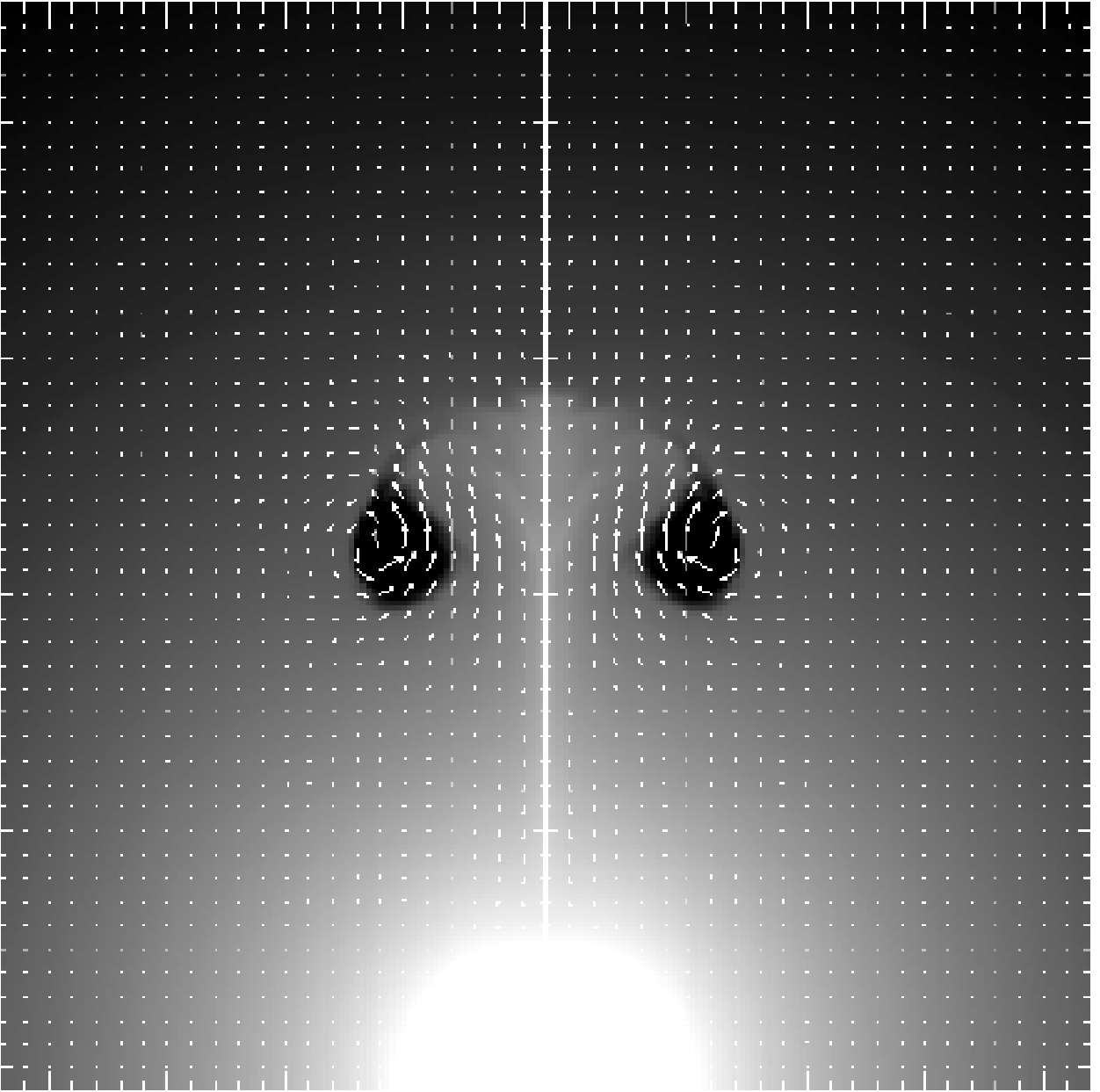}
\end{array}
$
\caption{ 
Flow velocity vectors at time $t=150t_0$
in the $\eta=10^{-3}$ 
simulation with triple the standard spatial resolution.
This is the 
approximate time when
the simulated radio morphology matches 
observations.
The maximum velocity is $\approx1.2v_0$.
The region and 
density scale are the same as in 
Figure~\ref{'fig.vfield000'}.
}
\label{'fig.vfield150'}
\end{center}
\end{figure}

\begin{figure}
\begin{center}
$
\begin{array}{c}
\plotone{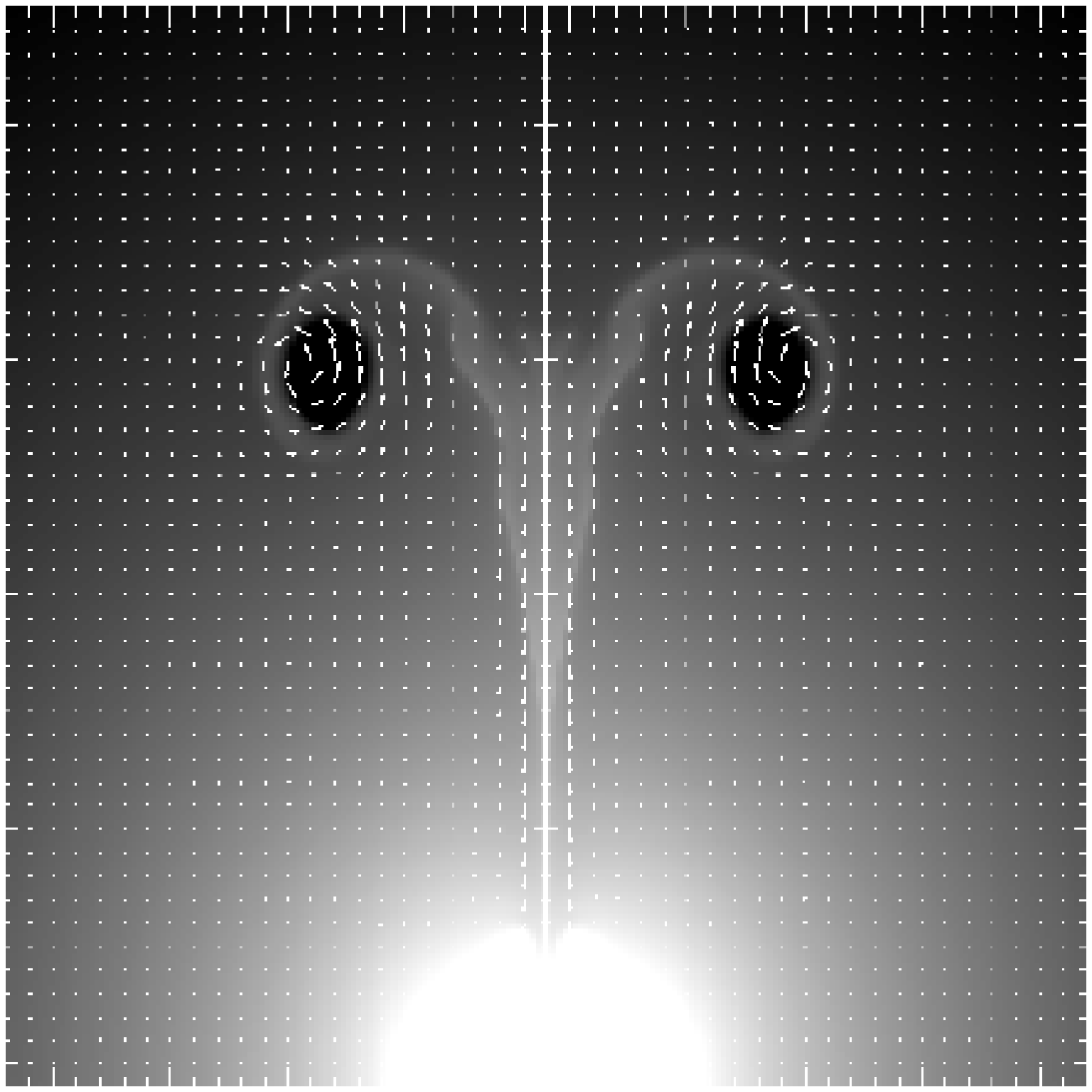}
\end{array}
$
\caption{ 
Flow velocity vectors at time $t=300t_0$ in the $\eta=10^{-3}$ simulation
with triple the standard spatial resolution.
The maximum velocity is $\approx1.0v_0$.
The region and density scale are the 
same as in Figure~\ref{'fig.vfield000'}.
}
\label{'fig.vfield300'}
\end{center}
\end{figure}

\begin{figure}
\begin{center}
$
\begin{array}{c}
\plotone{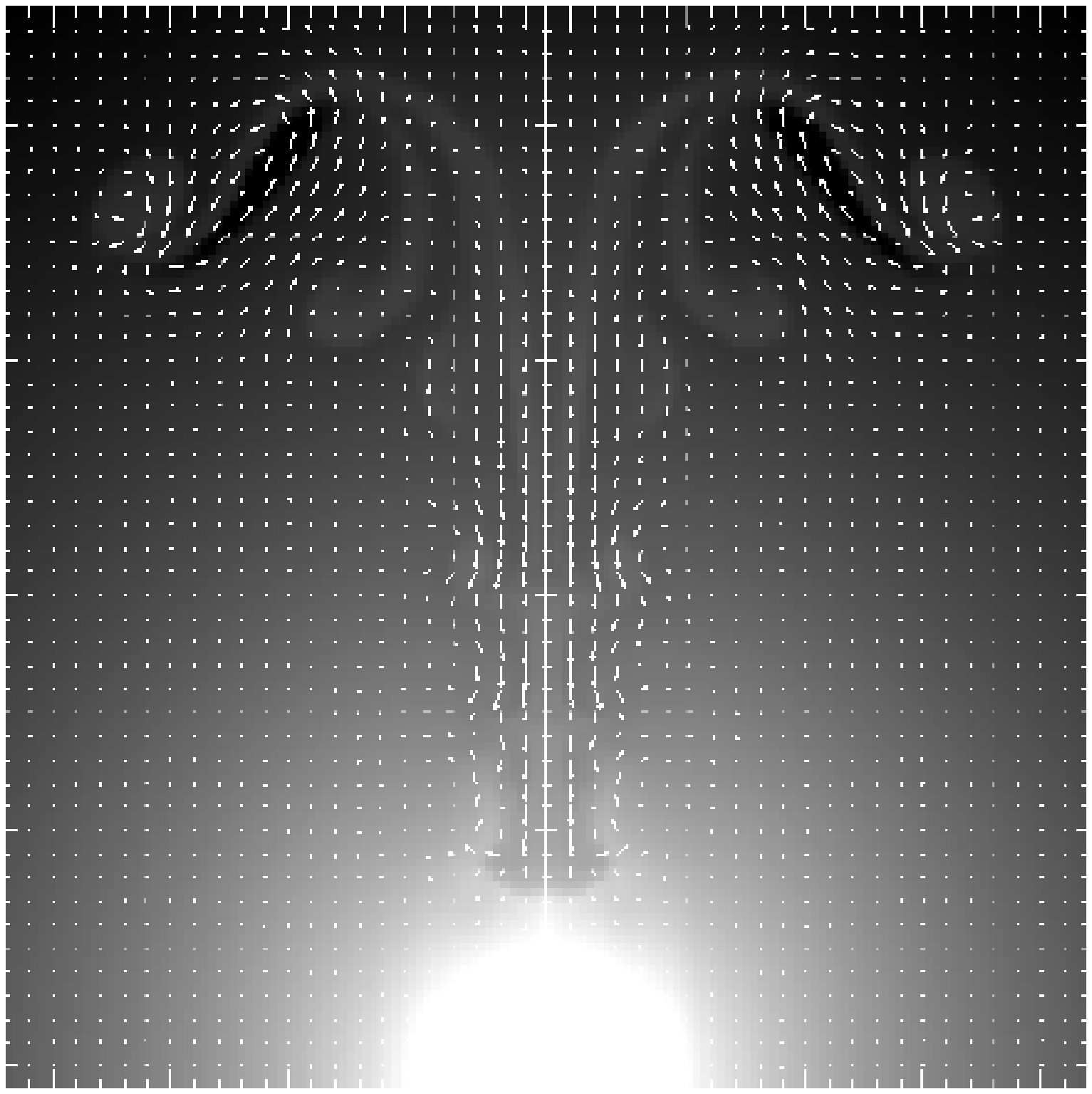}
\end{array}
$
\caption{ 
Flow velocity vectors at time $t=600t_0$ in the $\eta=10^{-3}$ simulation
with standard spatial resolution.
The maximum velocity is $\approx0.7v_0$.
The region and density scale are the same as in 
Figure~\ref{'fig.vfield000'}.
}
\label{'fig.vfield600'}
\end{center}
\end{figure}

\begin{figure}
\begin{center}
$
\begin{array}{c}
\plotone{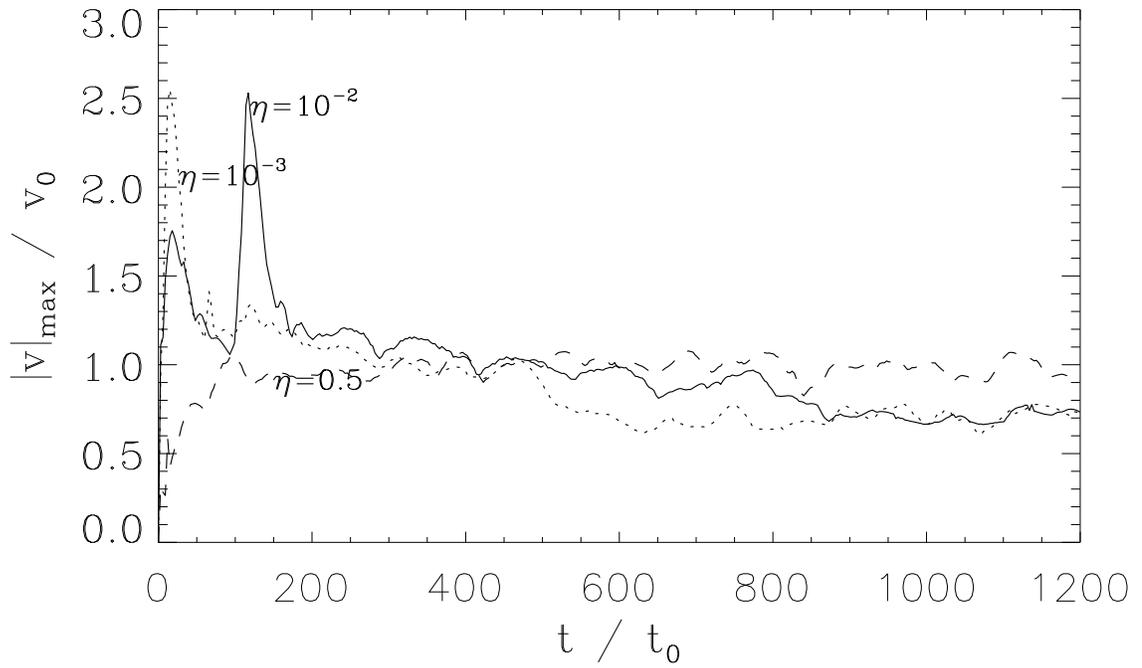}
\end{array}
$
\caption{ Maximum values of the velocity as a function of time, for cases
$\eta=0.5, 10^{-2}, 10^{-3}$ (dashed, solid and dotted lines
respectively).
The adiabatic sound speed in
the undisturbed gas is $\approx 1.29$.
}
\label{'fig.vmax.eta'}
\end{center}
\end{figure}

\begin{figure}
\begin{center}
$
\begin{array}{c}
\plotone{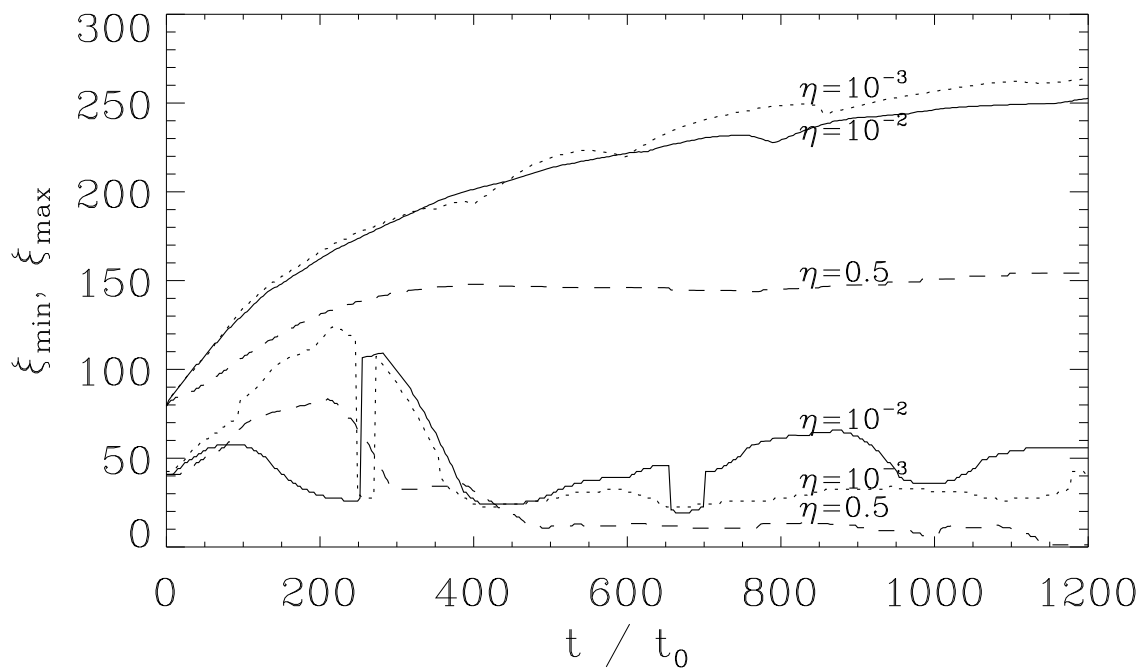}
\end{array}
$
\caption{ 
Maximum and minimum radial distances of bubble material from the galaxy nucleus,
$\xi_{\rm max}$ and $\xi_{\rm min}$,
as a functions of time for the cases with density contrasts
$\eta=0.5, 10^{-2}, 10^{-3}$ (dashed, solid and dotted respectively).
}
\label{'fig.xiveta'}
\end{center}
\end{figure}

\begin{figure}
\begin{center}
$
\begin{array}{c}
\plotone{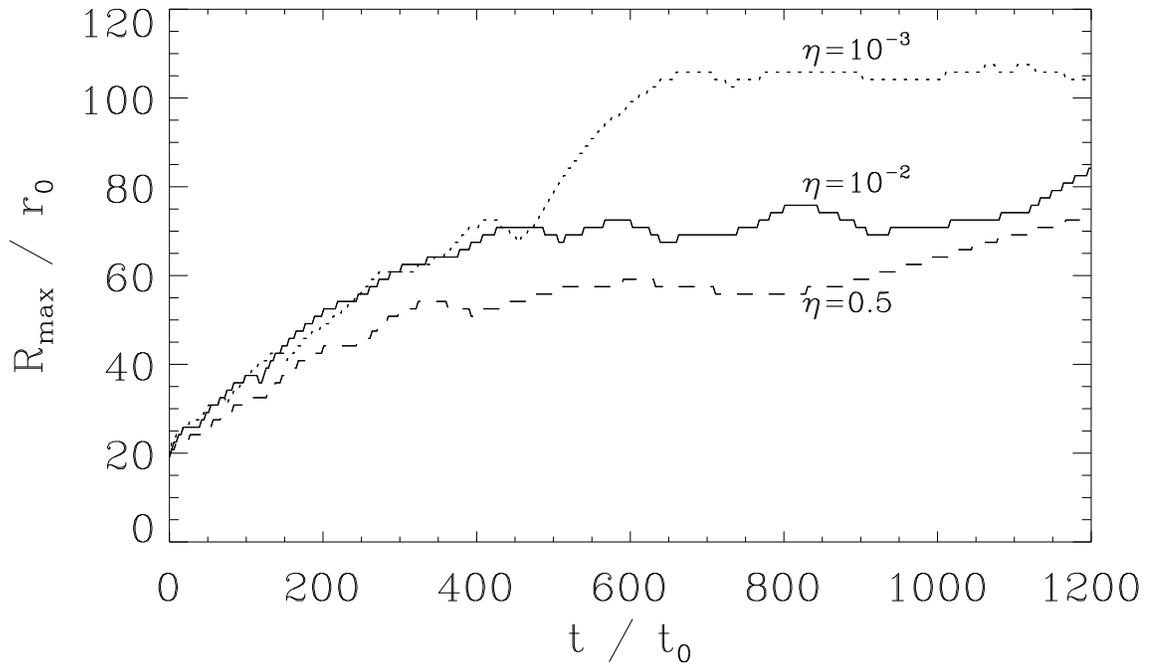}
\end{array}
$
\caption{ 
Extent of the bubble in the cylindrical radial direction 
$R_{\rm max}$ as a function of time
for the cases with density contrasts $\eta=0.5, 10^{-2}, 10^{-3}$.
}
\label{'fig.rveta'}
\end{center}
\end{figure}

\begin{figure}
\begin{center}
$
\begin{array}{c}
\plotone{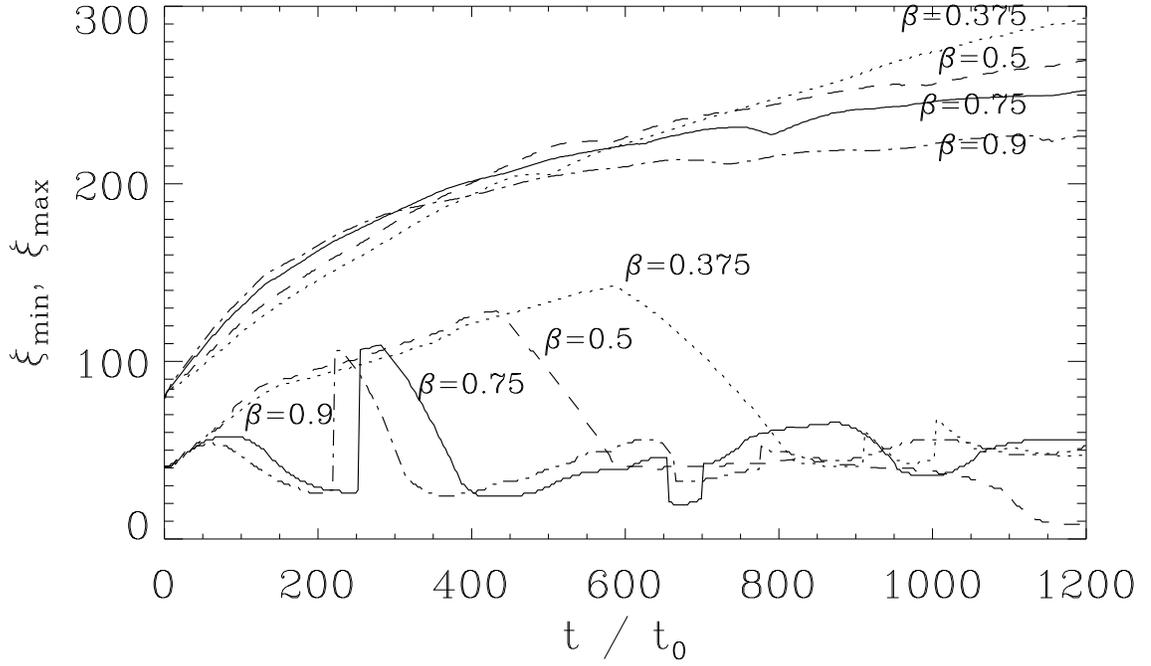}
\end{array}
$
\caption{ 
Maximum and minimum radial distances of bubble material from the galaxy nucleus,
$\xi_{\rm max}$ and $\xi_{\rm min}$,
as a functions of time for the cases with density contrast
$\eta=10^{-2}$ and
$\beta=0.375, 0.5, 0.75, 0.9$
(dotted, dashed, solid and dot-dashed lines respectively).
}
\label{'fig.xivbeta'}
\end{center}
\end{figure}

\begin{figure}
\begin{center}
$
\begin{array}{cc}
\psfig{file=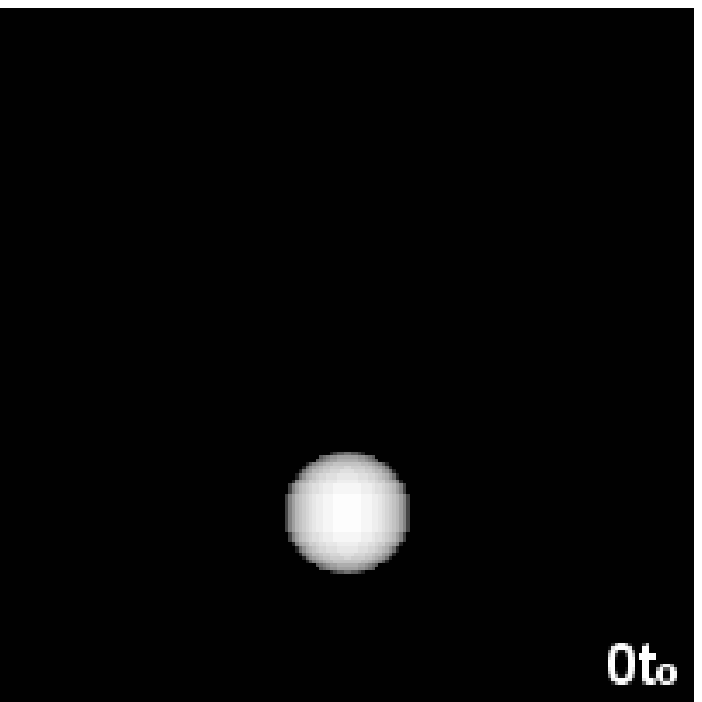,width=4cm}
& \psfig{file=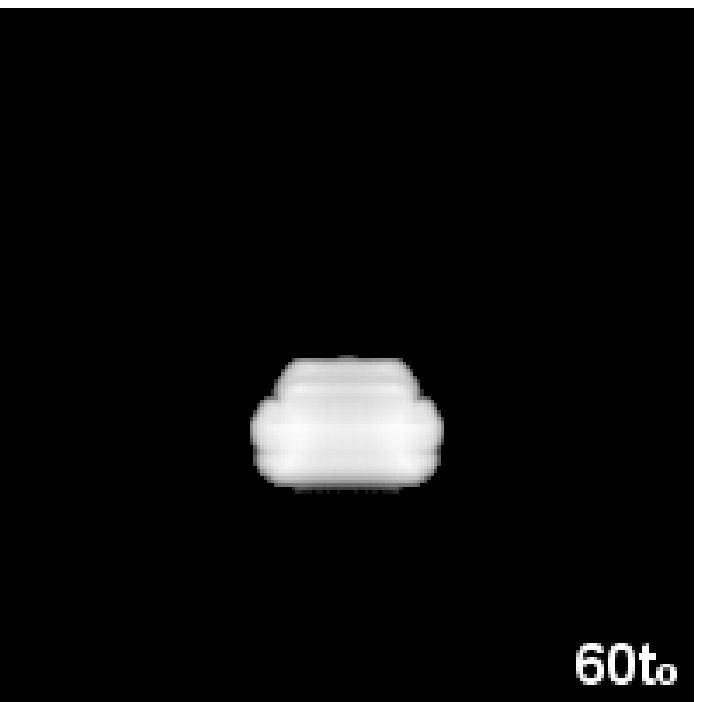,width=4cm}
\\
\psfig{file=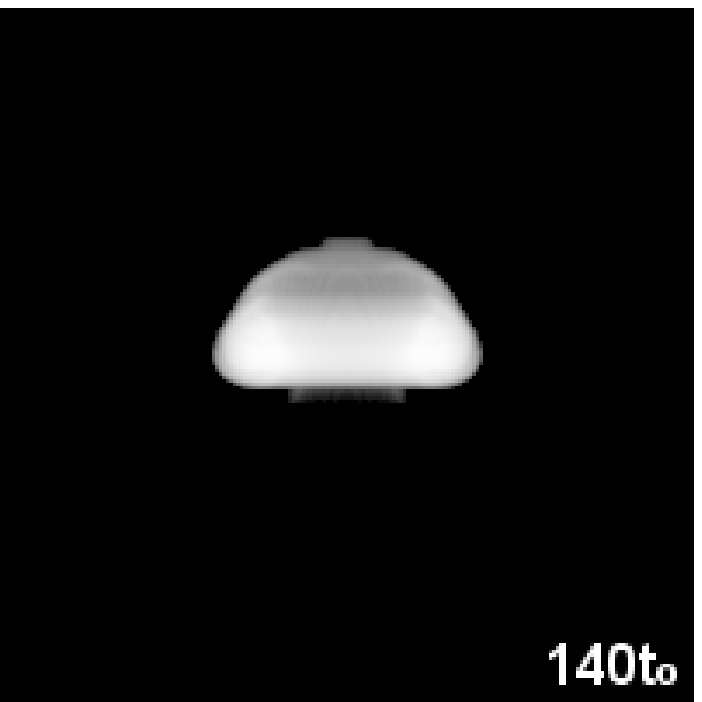,width=4cm} 
& \psfig{file=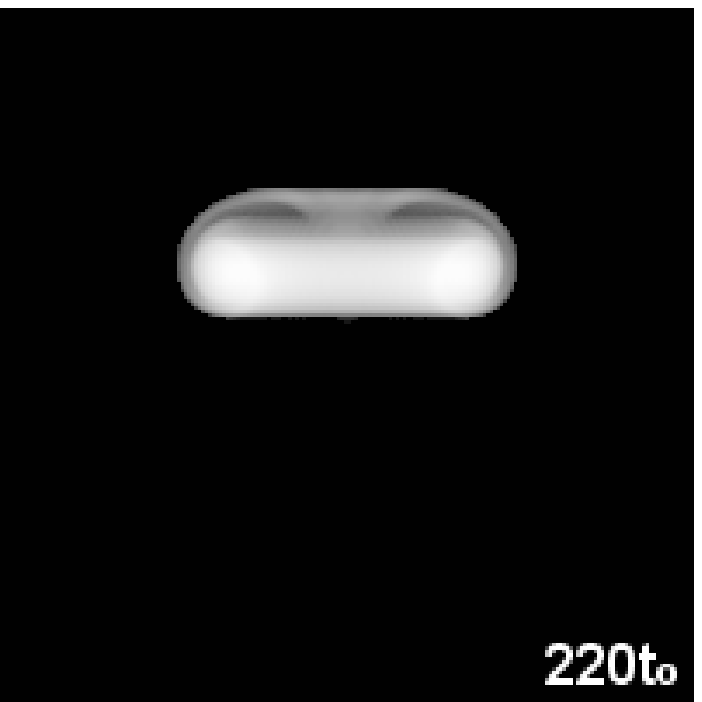,width=4cm}
\\
\psfig{file=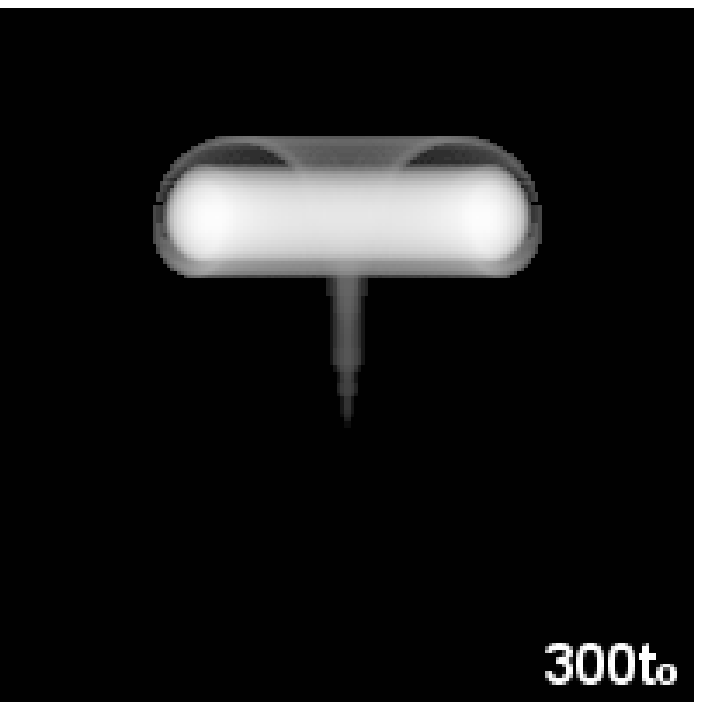,width=4cm} 
& \psfig{file=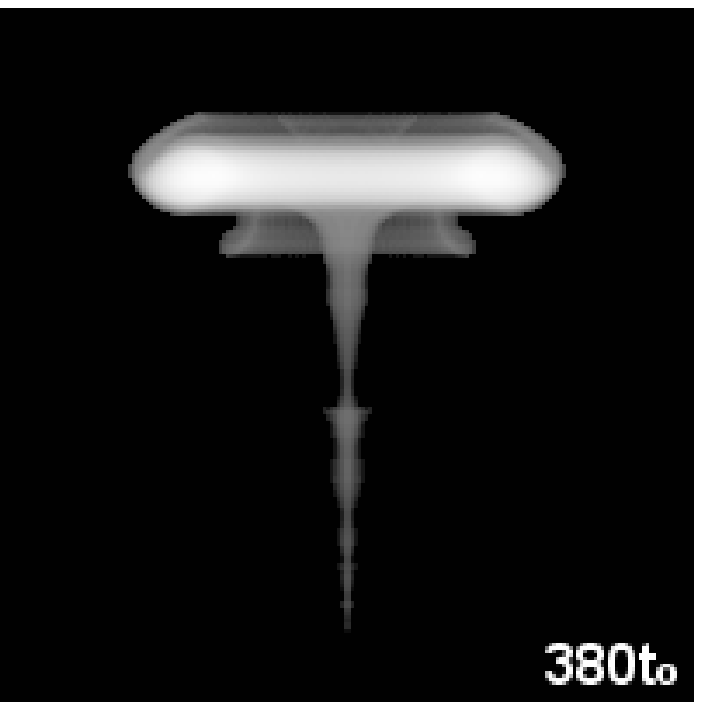,width=4cm}
%
\end{array}
$
\end{center}
\caption{ 
Raytraced simulated emission maps at different times.
The $z$ axis is vertical at right angles to the line of sight.
The origin is at the middle of the lower edge of each frame.
This is the case of $\eta=10^{-3}$ and $\beta=0.75$.
For the sake of clarity,
results were obtained at triple the spatial resolution
of our standard simulations.
The frame at $t=140t_0$ is in the ``young torus'' phase
with wispy emission above the main bubble torus.
We identify this stage with the observed radio morphology of the NML.
\label{'fig.hires'} 
}
\end{figure}

\begin{figure}
\begin{center}
$
\begin{array}{ccccccc}
\eta
\\
0.5
&\psfig{file=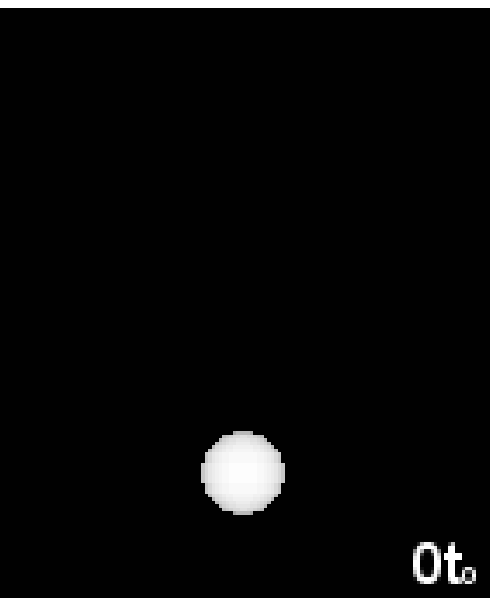,width=2.4cm}
&\psfig{file=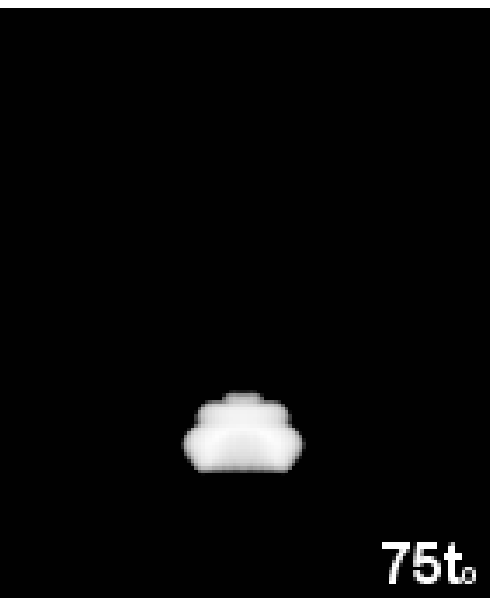,width=2.4cm}
&\psfig{file=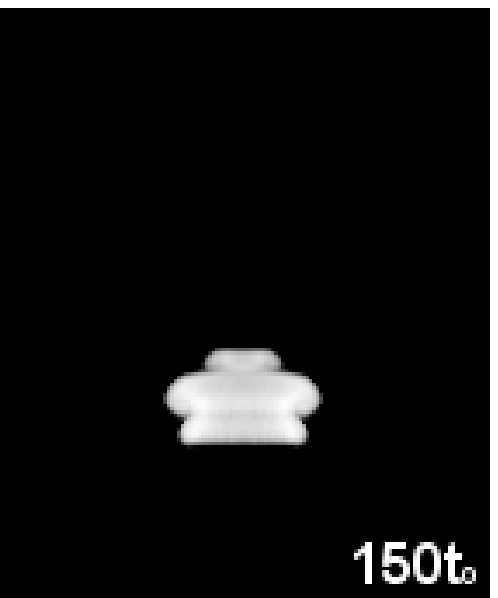,width=2.4cm}
&\psfig{file=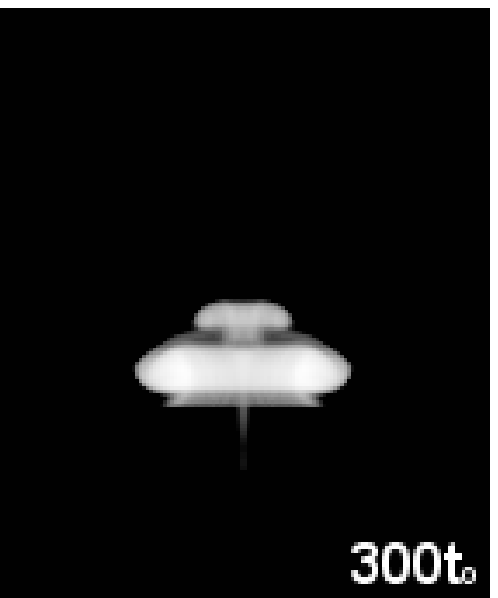,width=2.4cm}
&\psfig{file=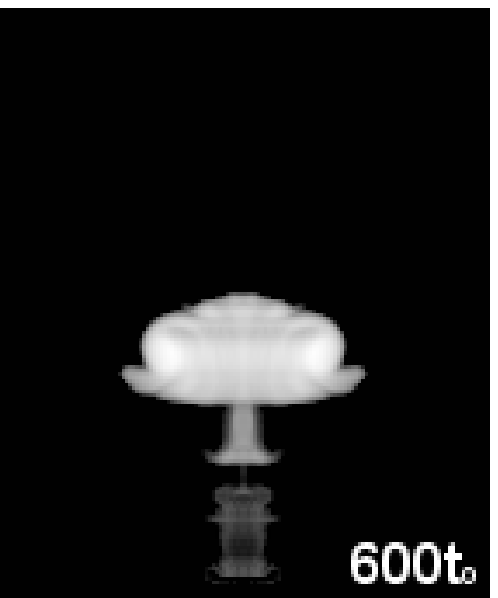,width=2.4cm}
&\psfig{file=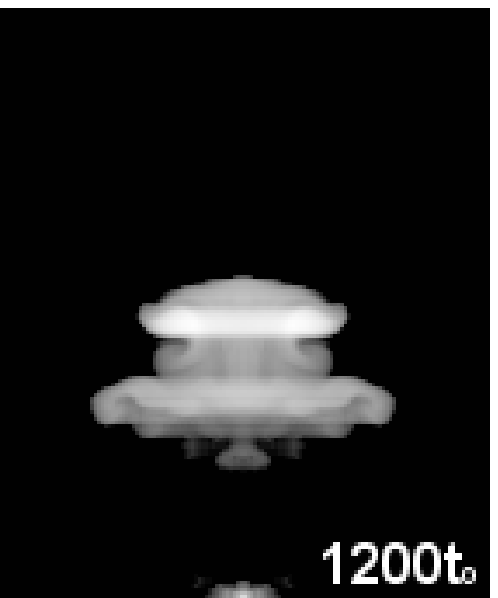,width=2.4cm}
\\
10^{-2}
&\psfig{file=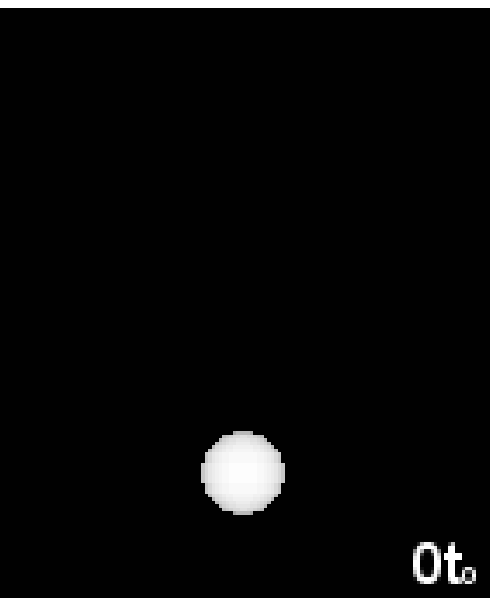,width=2.4cm}
&\psfig{file=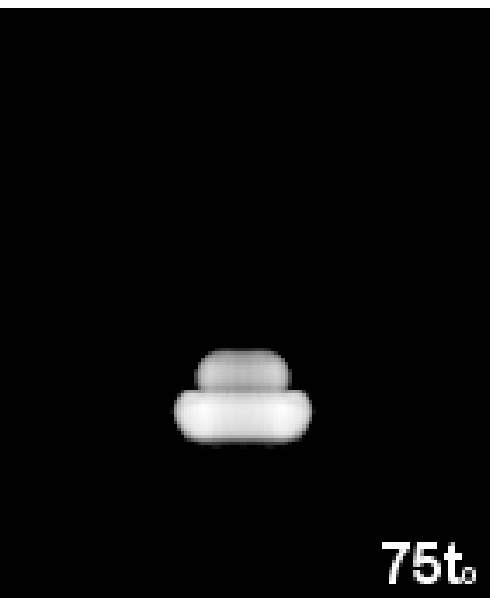,width=2.4cm}
&\psfig{file=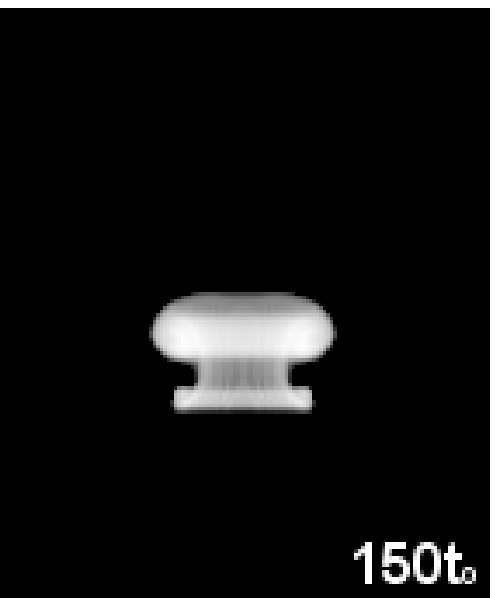,width=2.4cm}
&\psfig{file=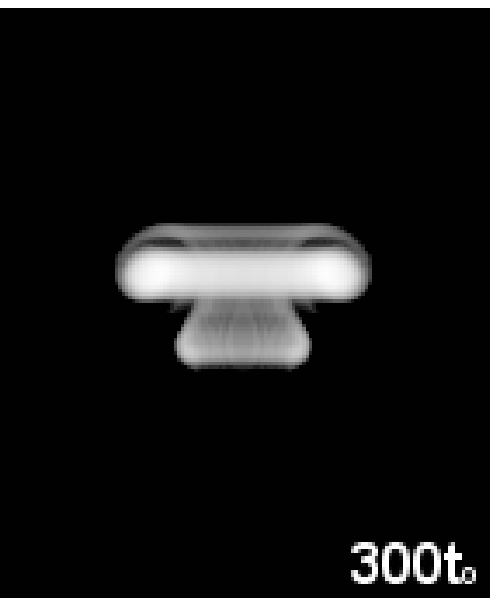,width=2.4cm}
&\psfig{file=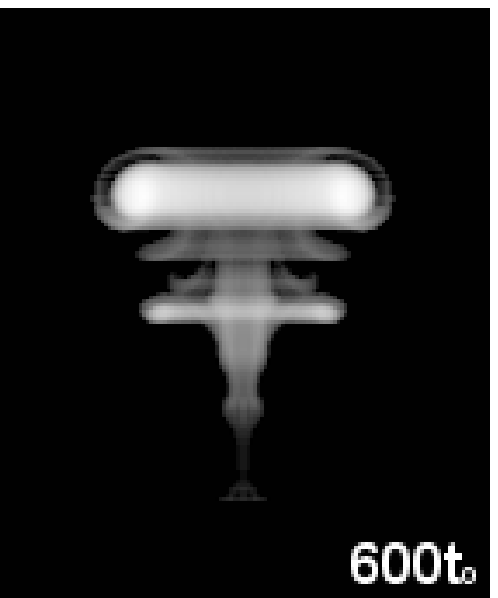,width=2.4cm}
&\psfig{file=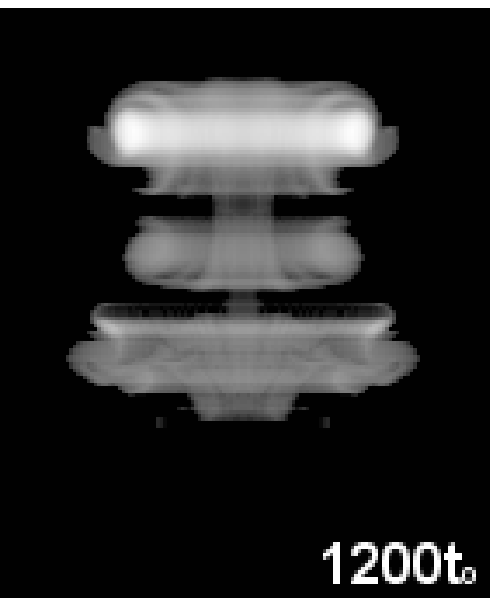,width=2.4cm}
\\
10^{-3}
&\psfig{file=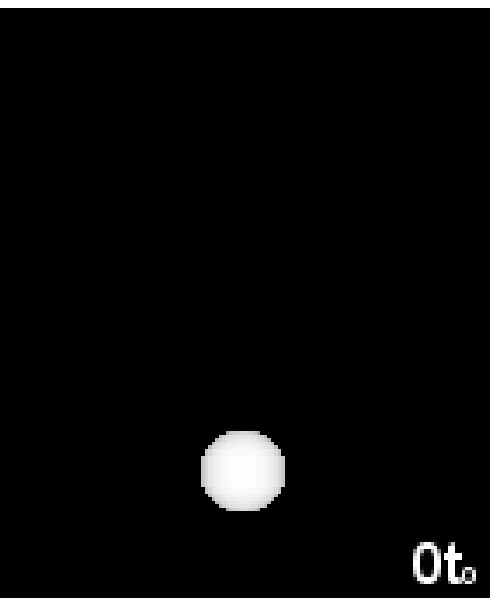,width=2.4cm}
&\psfig{file=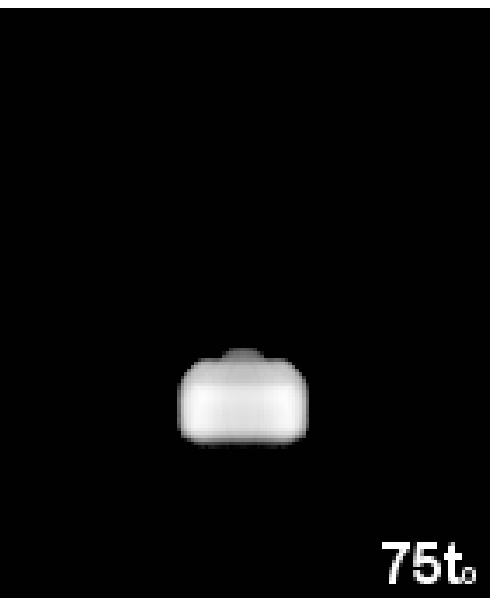,width=2.4cm}
&\psfig{file=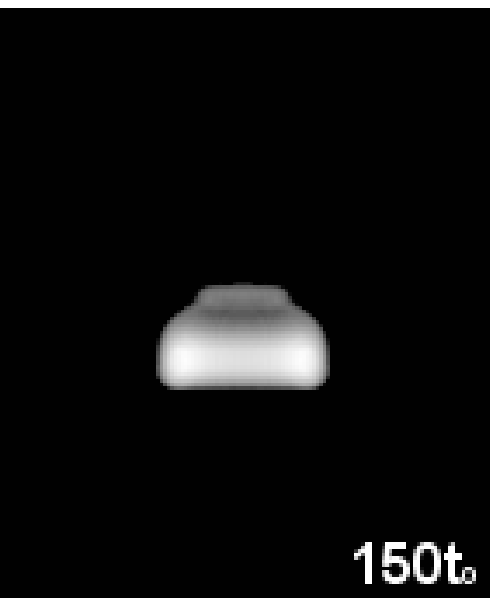,width=2.4cm}
&\psfig{file=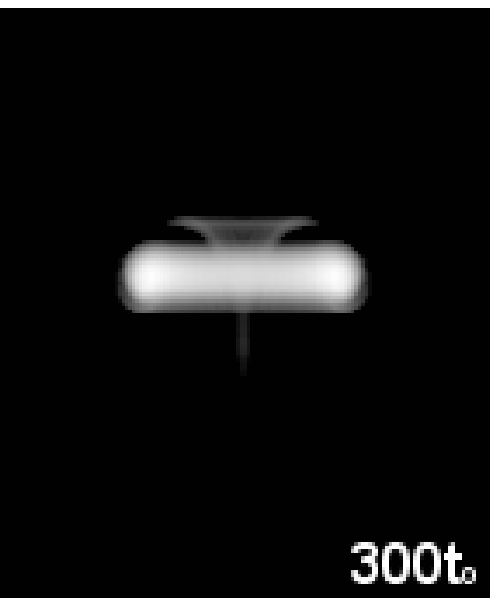,width=2.4cm}
&\psfig{file=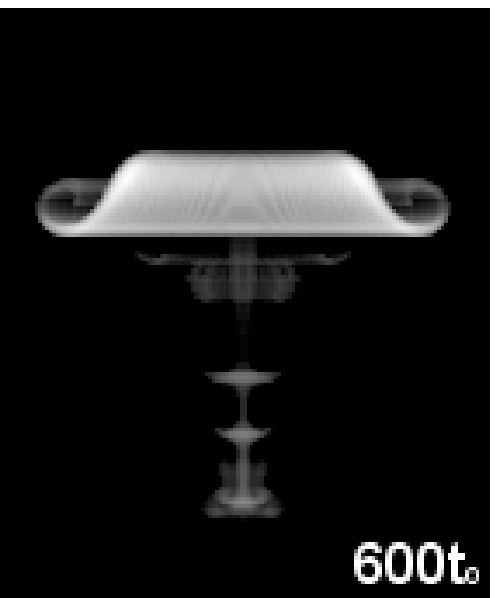,width=2.4cm}
&\psfig{file=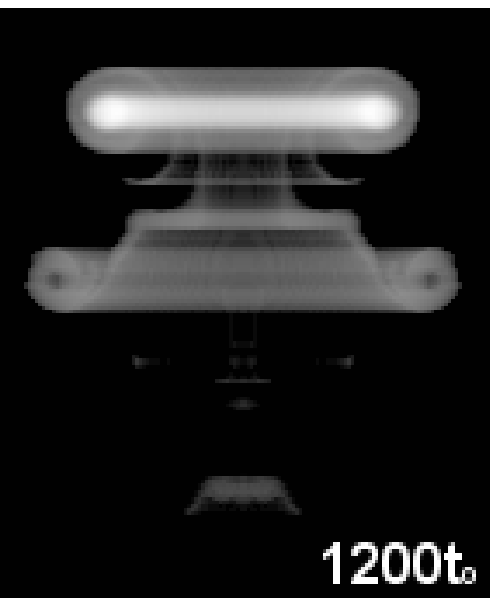,width=2.4cm}
\\ 
\end{array}
$
\end{center}
\caption{
Raytraced simulated emission maps at different times, 
comparing the morphologies of different cases of $\eta$, with $\beta=0.75$.
The axes are the same as in Figure~\ref{'fig.hires'}.
From top to bottom, the rows represent
the cases $\eta=0.5, 10^{-2}, 10^{-3}$ respectively.
}
\label{'fig.eta.morphologies'}
\end{figure}

\begin{figure}
\begin{center}
$
\begin{array}{ccc}
  \psfig{file=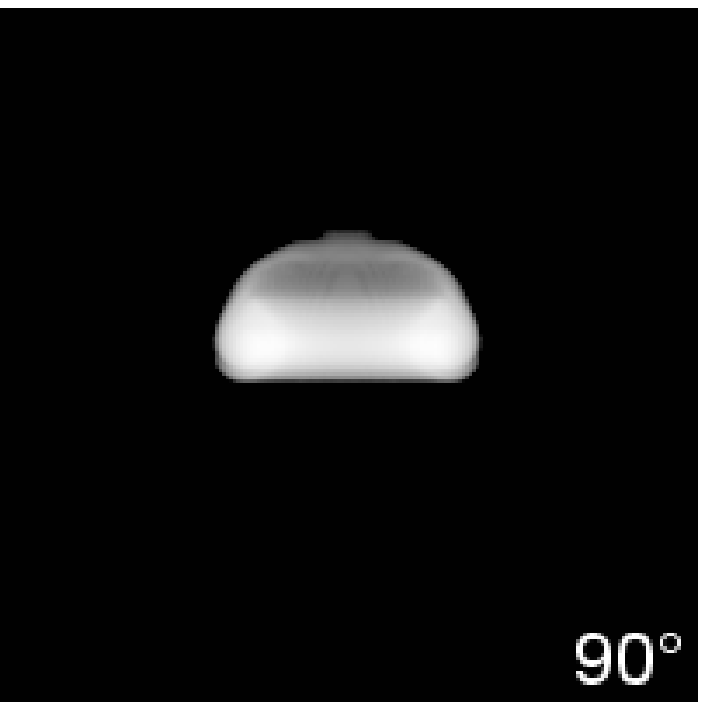,width=4cm}
& \psfig{file=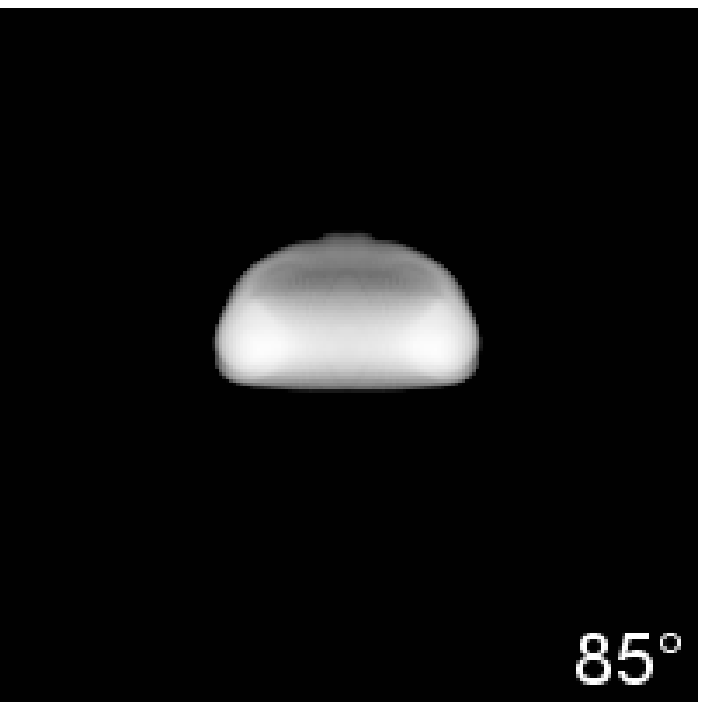,width=4cm}
& \psfig{file=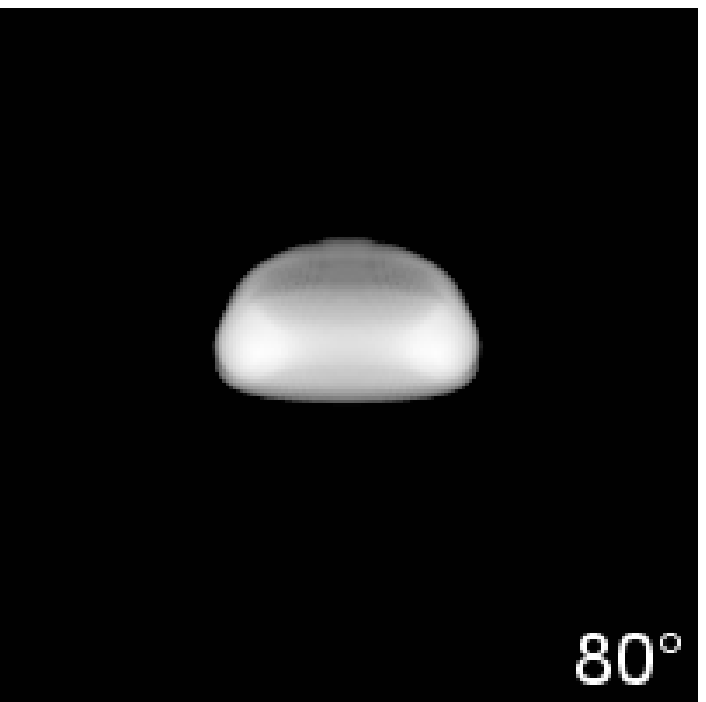,width=4cm}
\\
  \psfig{file=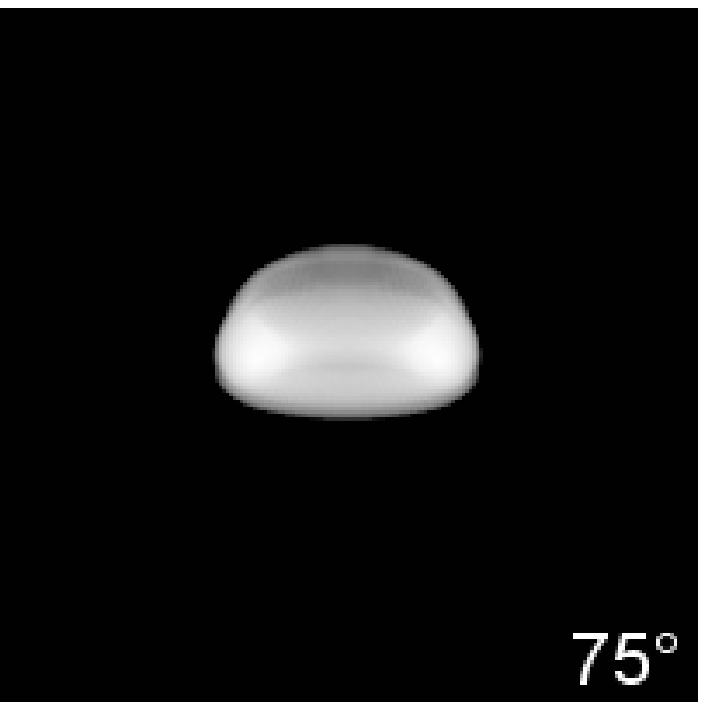,width=4cm}
& \psfig{file=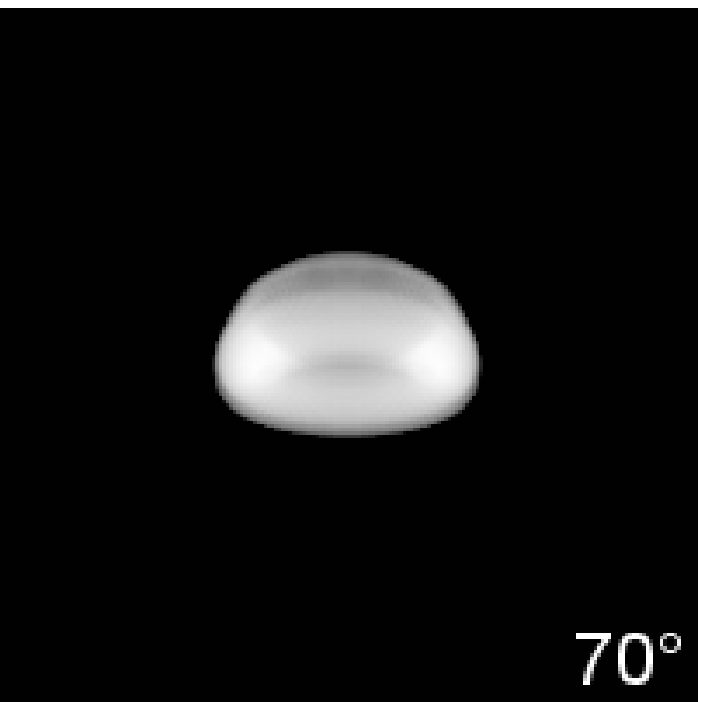,width=4cm}
& \psfig{file=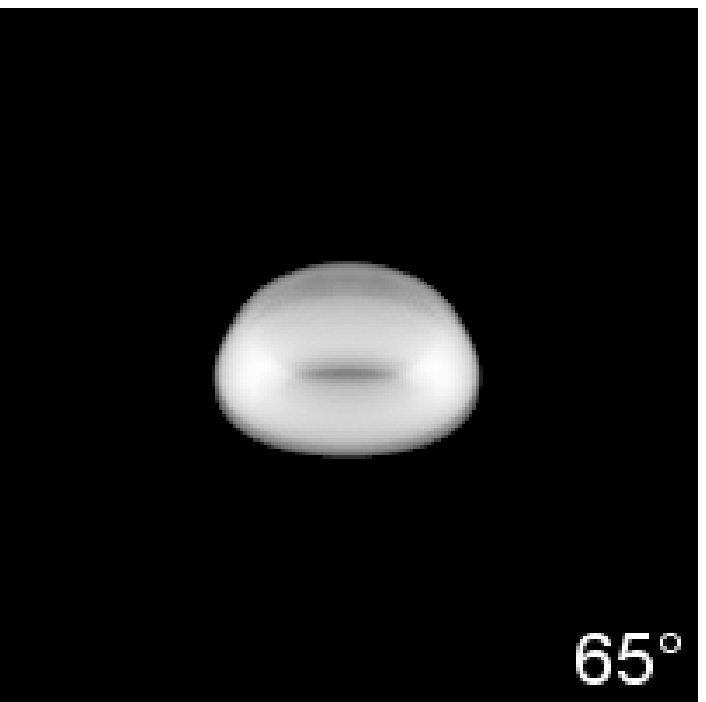,width=4cm}
\\
\end{array}
$
\end{center}
\caption{ 
Raytraced simulated emission maps at different inclinations 
for the same case as Figure~\ref{'fig.hires'},
\ie $\eta=10^{-3}$ and $\beta=0.75$, at time $t=150t_0$.
}
\label{'fig.inclination'}
\end{figure}